\documentclass[twocolumn, times, tighten]{aastex62}
\pdfoutput=1 
\usepackage{textcomp}
\usepackage{amsmath}
\usepackage{color}
\usepackage{amssymb}
\usepackage{hyperref}

\newcommand{\qvis}{$Q_{\mathrm{vis}}^{+}$}
\newcommand{\qrad}{$Q_{\mathrm{rad}}^{-}$}
\newcommand{\qill}{$Q_{\mathrm{mc}}^{+}$}

\shortauthors{Sun et al.}
\shorttitle{CHAR}

\begin{document}
\nocite{*}

\revised{\bf Draft: \today}
\title{Corona-Heated Accretion-disk Reprocessing (CHAR): A Physical Model to Decipher the Melody of AGN UV/optical Twinkling}

\author[0000-0002-0771-2153]{Mouyuan Sun}
\affiliation{Department of Astronomy, Xiamen University, Xiamen, Fujian 
361005, China; msun88@xmu.edu.cn}
\affiliation{CAS Key Laboratory for Research in Galaxies and Cosmology, 
Department of Astronomy, University of Science and Technology of China, 
Hefei 230026, China; xuey@ustc.edu.cn}
\affiliation{School of Astronomy and Space Science, University of Science 
and Technology of China, Hefei 230026, China}

\author[0000-0002-1935-8104]{Yongquan Xue}
\affiliation{CAS Key Laboratory for Research in Galaxies and Cosmology, 
Department of Astronomy, University of Science and Technology of China, 
Hefei 230026, China; xuey@ustc.edu.cn}
\affiliation{School of Astronomy and Space Science, University of Science 
and Technology of China, Hefei 230026, China}

\author[0000-0002-0167-2453]{W. N. Brandt}
\affiliation{Department of Astronomy \& Astrophysics, 525 Davey Lab, The Pennsylvania State
University, University Park, PA 16802, USA}
\affiliation{Institute for Gravitation and the Cosmos, 525 Davey Lab, The Pennsylvania State
University, University Park, PA 16802, USA}
\affiliation{Department of Physics, 104 Davey Lab, The Pennsylvania State University, University 
Park, PA 16802, USA}

\author[0000-0003-3137-1851]{Wei-Min Gu}
\affiliation{Department of Astronomy, Xiamen University, Xiamen, Fujian 
361005, China; msun88@xmu.edu.cn}

\author[0000-0002-1410-0470]{Jonathan R. Trump}
\affiliation{Department of Physics, University of Connecticut, Storrs, CT 06269, USA}

\author[0000-0002-4223-2198]{Zhenyi Cai}
\affiliation{CAS Key Laboratory for Research in Galaxies and Cosmology, 
Department of Astronomy, University of Science and Technology of China, 
Hefei 230026, China; xuey@ustc.edu.cn}
\affiliation{School of Astronomy and Space Science, University of Science 
and Technology of China, Hefei 230026, China}

\author[0000-0003-3667-1060]{Zhicheng He}
\affiliation{CAS Key Laboratory for Research in Galaxies and Cosmology, 
Department of Astronomy, University of Science and Technology of China, 
Hefei 230026, China; xuey@ustc.edu.cn}
\affiliation{School of Astronomy and Space Science, University of Science 
and Technology of China, Hefei 230026, China}

\author[0000-0003-1474-293X]{Da-bin Lin}
\affiliation{Guangxi Key Laboratory for Relativistic Astrophysics, Department of 
Physics, Guangxi University, Nanning 530004, China}

\author[0000-0001-8678-6291]{Tong Liu}
\affiliation{Department of Astronomy, Xiamen University, Xiamen, Fujian 
361005, China; msun88@xmu.edu.cn}

\author[0000-0002-4419-6434]{Junxian Wang}
\affiliation{CAS Key Laboratory for Research in Galaxies and Cosmology, 
Department of Astronomy, University of Science and Technology of China, 
Hefei 230026, China; xuey@ustc.edu.cn}
\affiliation{School of Astronomy and Space Science, University of Science 
and Technology of China, Hefei 230026, China}

\begin{abstract}
Active galactic nuclei (AGNs) have long been observed to ``twinkle'' (i.e., their brightness varies with 
time) on timescales from days to years in the UV/optical bands. Such AGN UV/optical variability is 
essential for probing the physics of supermassive black holes (SMBHs), the accretion disk, and the 
broad-line region. Here we show that the temperature fluctuations of an AGN accretion disk, which 
is magnetically coupled with the corona, can account for observed high-quality AGN optical light 
curves. We calculate the temperature fluctuations by considering the gas physics of the accreted 
matter near the SMBH. We find that the resulting simulated AGN UV/optical light curves share the 
same statistical properties as the observed ones as long as the dimensionless viscosity parameter 
$\alpha$, which is widely believed to be controlled by magnetohydrodynamic (MHD) turbulence in 
the accretion disk, is about $0.01$---$0.2$. Moreover, our model can simultaneously explain the 
larger-than-expected accretion disk sizes and the dependence of UV/optical variability upon 
wavelength for NGC 5548. Our model also has the potential to explain some other observational 
facts of AGN UV/optical variability, including the timescale-dependent bluer-when-brighter color 
variability and the dependence of UV/optical variability on AGN luminosity and black 
hole mass. Our results also demonstrate a promising way to infer the black-hole mass, the accretion 
rate, and the radiative efficiency, thereby facilitating understanding of the gas physics and MHD 
turbulence near the SMBH and its cosmic mass growth history by fitting the AGN UV/optical light 
curves in the era of time-domain astronomy. 
\end{abstract}

\keywords{accretion, accretion disks---galaxies: active---quasars: general---quasars: 
supermassive black holes}

\section{Introduction}
\label{sect:intro}
The ultra-violet (UV) to optical continuum emission of active galactic nuclei (AGNs) 
is widely believed to be emitted by a geometrically thin but optically thick accretion 
disk \citep[i.e., the classical standard thin disk, hereafter SSD; see, e.g.,][]{SSD, 
Czerny1987}. The gravitational energy released in the disk is balanced by the 
blackbody radiative cooling, and the effective temperature decreases with 
increasing distance from the central supermassive black hole (SMBH). The UV-to-optical 
emission is a superposition of multi-temperature blackbody radiation. The 
expected UV-to-optical spectral energy distribution (SED), however, might be altered 
by additional physical processes, e.g., strong disk winds \citep[e.g.,][]{Slone2012, 
Laor2014, Li2019, Sun2019} or a disk atmosphere \citep[e.g.,][]{Hall2018}. Also, 
for very faint or luminous AGNs, cooling due to advection or photon trapping plays 
an important role \citep[e.g.,][]{slim, adaf}. In the innermost regions or above the 
accretion disk, there also exists a hot and optically thin corona which produces 
hard X-ray emission \citep[e.g.,][]{Haardt1991, Liu2002}. 

UV and optical emission often possesses small-amplitude ($\sim 10\%$ on timescales 
of a few years) stochastic variability; violent AGN flares are also 
observed in a small fraction of AGNs \citep[e.g., ][]{MacLeod2016, Yang2018}. 
The statistical properties of AGN UV and optical stochastic variations have been 
explored in great detail in many works. The major observational results of these 
works can be summarized as follows. 
\begin{enumerate}
\item  A damped random walk (DRW) process (whose power spectral density, 
PSD, $P(f) \propto 1/(f_0^2+f^2)$, where $f_0=1/\tau$ is the damping frequency) 
seems to be able to describe AGN UV/optical variability on timescales of months to 
years \citep{Kelly2009, MacLeod2010, Zu2013}. On very short timescales (e.g., 
days), the observed variability amplitude is lower than the prediction of the DRW 
model \citep{Mushotzky2011, Kasliwal2015, Smith2018}. On very long timescales 
(e.g., several decades), the DRW model seems to under-predict the actual variability 
amplitude \citep{MacLeod2012, Guo2017}. 

\item AGN UV/optical fractional variability amplitude increases with decreasing rest-frame 
wavelength \citep[i.e., UV emission is more variable than optical emission; see, 
e.g.,][]{MacLeod2010, Morganson2014, Sun2015, Simm2016, Sanchez2018}. 

\item AGN UV/optical fractional variability amplitude anti-correlates with AGN luminosity 
\citep[e.g.,][]{MacLeod2010, Zuo2012, Morganson2014, Li2018, Sun2018c}, the iron strength 
\citep[i.e., the ratio of optical iron emission to H$\beta$; see, e.g.,][]{Ai2010, Sun2018c}, 
Eddington ratio \citep[e.g.,][]{MacLeod2010, Zuo2012, Simm2016}, or additional 
parameters \citep[e.g.,][]{Kang2018}. 
 
\item The damping timescale $\tau$ correlates with AGN luminosity \citep{Sun2018c}, 
black-hole mass ($M_{\mathrm{BH}}$), or wavelength \citep{MacLeod2010}. 

\item AGN color tends to follow a bluer-when-brighter pattern \citep[e.g.,][]{Ruan2014}. 
The bluer-when-brighter behavior seems to be more evident on timescales 
of weeks to months rather than on timescales of years \citep{SunYH2014}. 

\item Variations in different bands are well coordinated. Changes of short-wavelength 
emission lead those of long-wavelength emission \citep[e.g.,][]{Sergeev2005, 
Fausnaugh2016, Jiang2017, Homayouni2018, Cackett2018, Kokubo2018, McHardy2018, 
Mudd2018, Yu2018, Edelson2019}. Current observations have a broad diversity of 
measured time lags beyond the SSD theory: some AGNs have time lags that are about 
three times larger than the flux-weighted light-travel time delays of the SSD theory. The 
time lags between X-ray and UV emission can be even about ten times larger than the 
expectations of the SSD theory, and their correlations often seem to be weak 
\citep{Edelson2019}. 

\item AGN microlensing observations also suggest that the accretion-disk sizes are 
larger than the flux-weighted radii of the SSD theory \citep[e.g.,][]{Morgan2010, 
Cornachione2019}. 
\end{enumerate}

Variations in the UV-to-optical bands might be caused by the reprocessing of variable X-ray 
emission \citep{Collin1991, Krolik1991}. In the X-ray reprocessing model, as the variable X-ray 
emission propagates to the disk surface, it is absorbed by the cold disk surface. It is then 
reprocessed immediately in the UV/optical bands (see Eq.~\ref{eq:qqq} and Section~\ref{sect:setup}). 
That is, UV/optical emission is expected to vary in response to X-ray light curves after 
a light-travel time delay; the time delay increases with increasing wavelength since the 
effective temperature anti-correlates with radius. The UV/optical inter-band correlations and 
time lags are indeed observed (see the observational fact \# 6). They can be used to probe the 
temperature profile and to constrain the fundamental physical processes of the accretion disk 
\citep[e.g.,][]{Lawrence2018}. However, the expected tight correlations between X-ray and UV/optical 
emission are not observed at least for some AGNs \citep[for a summary, see][]{Edelson2019}. 

AGN broad emission lines (BELs) arise from Doppler-broadened line emission from gas clouds 
in the broad line region (BLR); these BLR gas clouds are photoionized by the extreme UV 
(EUV) emission. BELs are also expected to respond to EUV emission after a light-travel time 
delay; the time delay can probe the spatial structure of BLR \citep[i.e., Reverberation 
Mapping, hereafter RM;][]{Blandford1982}. The EUV variations are not monitored for most 
RM AGNs; instead, the time lags between BELs and the nearby UV/optical continua are 
measured. The underlying assumptions are the following: first, the EUV and the nearby 
UV/optical emission is tightly correlated; second, the time lags between EUV and the 
nearby UV/optical emission are negligible with respect to the time delays of BELs. The 
first assumption is probably robust since the tight correlations between BELs and the nearby 
UV/optical continua are indeed observed \citep[for exceptions, see, e.g.,][]{Goad2016}. 
In addition, a good correlation between the EUV and the $1350\ \mathrm{\AA}$ UV emission 
indeed exists for NGC 5548 \citep[see Figures 3 \& 4 of ][]{Marshall1997}. The second 
assumption may require some attention \citep[e.g.,][]{Vestergaard2019} since the inter-band time 
lags are larger than the expectations of the SSD theory for at least some AGNs (see 
observational fact \# 6). Nevertheless, the distance of the BLR to the central SMBH was 
measured for some AGNs with diverse properties \citep[e.g.,][]{Bentz2009, Du2016, Grier2017, 
Grier2019}, which enables us to estimate $M_{\mathrm{BH}}$ of non-local 
SMBHs \citep[for recent reviews, see, e.g.,][]{Shen2013, Peterson2014}. Therefore, exploring 
AGN UV/optical variability is of fundamental importance to our understanding of black-hole 
accretion and our improvement of $M_{\mathrm{BH}}$ measurements. 

In the era of time-domain astronomy, large time-domain surveys like LSST \citep{Ivezic2019} 
will offer a tremendous amount of AGN variability data \citep[e.g.,][]{Brandt2018}. These data 
can refine the observational conclusions summarized above, help constrain accretion disk 
models, and obtain AGN physical parameters. These goals can be achieved if we correctly 
understand the physical origin of AGN UV/optical variability \citep[rather than adopting 
more complicated empirical stochastic models;][]{Vio2005}. 

A few models have been proposed to explain AGN UV/optical variability \citep[for a brief 
discussion, see, e.g.,][]{Czerny2004, Czerny2006}. For instance, global \citep{Li2008, Liu2016} 
or local \citep{Lyubarskii1997} accretion-rate fluctuations might induce the observed UV/optical 
variations. However, the timescale for the accretion rate to change is the viscous timescale, 
which is $\gtrsim 100$ yr for the UV/optical emission regions; therefore, this timescale is much 
longer than our current observations. 
Instead, local temperature fluctuations \citep{Kelly2009}, which should happen on a much 
shorter timescale, the thermal timescale (see also Eq.~\ref{eq:scale}), are suggested to be 
responsible for UV/optical variability. The temperature fluctuations are often \textit{assumed} 
to follow the DRW process \citep[e.g.,][]{Dexter2011}. Such a temperature fluctuation 
model \citep[with further modifications; see][]{Cai2016} has the potential to explain the 
bluer-when-brighter tendency and its timescale dependence. However, this model cannot 
explain the inter-band correlations \citep[e.g.,][]{Kokubo2015}. \cite{Cai2018} suggested that 
there are fast-propagating temperature fluctuations (possibly 
caused by strong outflows or variability in corona; the detailed physical mechanism remains 
unclear) in the accretion disk. By again assuming that the resulting temperature fluctuations 
follow the DRW process, they constructed a model to explain the larger-than-expected 
inter-band time lags in several nearby AGNs. The assumption of DRW-fluctuations in 
\cite{Dexter2011} and \cite{Cai2018} is mostly motivated by observations rather than directly 
by the gas physics of matter near the central SMBH. 

The timescale problem can also be avoided by considering X-ray reprocessing because 
of the following reasons. First, the X-ray emission regions are expected to be compact, and the 
relevant timescales should be very short. Second, the (variable) hard X-ray photons should be 
absorbed in the surface layer of the thin cold accretion disk, and the corresponding response 
timescale is extremely small \citep{Czerny2006}. However, 
the expected inter-band time lags are too small to be consistent with observations (see 
observational fact \# 6) although this discrepancy could be resolved by adding an additional 
ingredient, e.g., non-blackbody emission \citep{Hall2018} or strong winds \citep{Sun2019}. 
Moreover, the simplest X-ray reprocessing model also predicts too much short-term variability 
\citep[note that this inconsistency can be solved by replacing the X-ray corona with a UV 
torus;][]{Gardner2017}.  There are additional fundamental observational challenges. According 
to the simplest X-ray reprocessing model, the light curves at all different wavelengths should 
vary in a very similar way. This prediction is inconsistent with the observational facts 
\# 2, 4, and 5 \citep{Zhu2018}. Meanwhile, as mentioned above, the expected tight correlations 
between X-ray and UV/optical emission are not observed at least for some AGNs \citep[for a 
summary, see][]{Edelson2019}. Last but not least, there is a long-standing energy-budget
problem \citep[e.g.,][]{Clavel1992,Dexter2019} in the X-ray reprocessing model. According 
to this model, the X-ray luminosity should be comparable to the internal dissipation rate 
to produce the observed fractional variability of $10\%\sim 30\%$ in the UV/optical bands. At least for 
luminous AGNs, the required X-ray luminosity is likely to be too large to be consistent with X-ray 
observations \citep[e.g.,][]{Just2007}. The energy-budget problem is even more serious for 
highly variable AGNs \citep[i.e., AGNs with UV/optical fractional variability amplitudes of about several 
to ten; see, e.g.,][]{Dexter2019}. Therefore, the X-ray reprocessing model is unlikely to fully drive AGN 
UV/optical variability. 

In this work, we propose a new model, i.e., Corona-Heated Accretion-disk Reprocessing 
(hereafter CHAR), to understand AGN UV and optical variability. In this model, 
the outer ($\gtrsim 10$ Schwarzschild radius) disk and the innermost corona are efficiently 
coupled via the \textit{magnetic field}. As the magnetic turbulence/flaring in the corona drives 
X-ray variability, the same process also changes the accretion-disk heating rate and 
induces its temperature fluctuations (see Section~\ref{sect:model}). The energy-budget 
problem mentioned above might be avoided in our magnetic coupling picture if the corona is 
radiatively inefficient where most energy is carried by 
protons rather than electrons, and protons and electrons are decoupled (see e.g., 
\citealt{DiMatteo1998, Rozanska2000}). If so, only a small fraction of the power of the magnetic 
flares/turbulence in the corona drives X-ray emission, the remaining of which can affect the disk 
interior and induce significant UV/optical variability. In our CHAR model, we can determine 
the statistical properties of temperature fluctuations and AGN UV/optical light curves (the 
correlation between X-ray and UV/optical variability is briefly discussed in Sections~\ref{sect:ccf} 
and \ref{sect:dis2}) by considering the thermal-energy conservation law of an AGN 
accretion disk. 

This paper is formatted as follows. In Section~\ref{sect:model}, we present our model. 
In Section~\ref{sect:result_all}, we demonstrate the results of our CHAR model. 
In Section~\ref{sect:test}, we apply our model to explain high-quality Kepler AGN 
light curves and the multi-wavelength light curves and inter-band time lags of NGC 5548.
In Section~\ref{sect:dis}, we discuss the assumptions of our CHAR model, compare 
our CHAR model with some previous works, and forecast AGN UV/optical variability 
in the era of time-domain astronomy. Our conclusions are summarized in 
Section~\ref{sect:sum}. The Schwarzschild radius $R_{\mathrm{S}}\equiv 
2GM_{\mathrm{BH}}/c^2$, where $G$ and $c$ are the gravitational constant and 
speed of light, respectively. We adopt a flat $\Lambda$CDM cosmology with $h_0=0.7$ 
and $\Omega_{\rm{M}} = 0.3$.

\section{The Model}
\label{sect:model}
\subsection{Model Setup}
\label{sect:setup}
The outer parts (i.e., $R\gg R_{\mathrm{S}}$) of an accretion disk might receive external 
illumination via X-ray emission from a hot corona or the FUV emission from the inner 
(e.g., within $\sim 10 R_{\mathrm{S}}$) disk \citep[e.g.,][]{Gardner2017}. A significant 
fraction of the illuminating emission will be absorbed by the thin surface of the outer 
accretion disk. In the simplest 
X-ray reprocessing model \citep[e.g.,][]{Starkey2017}, it is often assumed that the 
absorbed energy is fully re-radiated away locally, i.e., the radiation cooling rate per surface 
area\footnote{Throughout this work, the heating/cooling rate are always per surface area, 
unless otherwise specified.} satisfies the following relation, 
\begin{equation}
\label{eq:qqq}
 Q_{\mathrm{rad}}^{-}(t)=2\sigma T^4_{\mathrm{eff}}= Q_{\mathrm{vis}}^{+}(t)+ 
 Q_{\mathrm{X}}^{+}(t-R_X/c) \\ ,
 \end{equation}
where $\sigma$, $T_{\mathrm{eff}}$, $R_X/c$, and $Q_{\mathrm{X}}$ denote the Stefan-Boltzmann 
constant, the effective temperature, the light-travel time lag between the hot corona/the innermost 
disk and the outer disk, and the external heating rate due to the X-ray corona, respectively; 
$R_X= \sqrt{H^2 +R^2}$ is the distance to the corona, where $H$ and $R$ are the distance of the corona 
and disk with respect to the SMBH, respectively. In the lamppost approximation \citep{Cackett2007}, 
$Q_{\mathrm{X}}=(1-A)L_{\mathrm{X}}H/(2\pi R^3)$ where $A$ is the albedo of the disk surface. 
$L_{\mathrm{X}}$ can vary on timescales of days or less since the X-ray external emission is produced in 
very compact regions (e.g., within $\sim 10 R_{\mathrm{S}}$), and various MHD instabilities 
may occur \citep{Noble2009}. If we neglect possible fluctuations in \qvis\ , 
\qrad\ should vary in response to $L_{\mathrm{X}}$ after a light-travel time lag. However, this 
simple model fails to explain many observational facts (see Section~\ref{sect:intro}). 

Magnetic fields play a fundamental role in accretion-disk theories since the magnetorotational 
instability (MRI) is widely believed to be responsible for producing viscosity in the accretion 
disk and converting gravitational energy into MHD turbulence. Then, the MHD turbulence 
dissipates and transfers the magnetic energy to heat the gas in the accretion 
disk, which produces the observed multi-wavelength emission. In the classical $\alpha$-prescription 
of viscosity \citep[e.g.,][]{SSD}, it is assumed that MHD turbulence is controlled by the total 
pressure in the accretion disk. However, recent numerical MHD simulations of accretion disks 
reveal the opposite behavior, i.e., MHD turbulence controls heat fluctuations in the accretion 
disk \citep[e.g.,][]{Hirose2009, Jiang2013} on timescales from the local orbital timescale to 
hundreds of times of the thermal 
timescale.\footnote{A delayed $\alpha$-prescription is proposed by two independent 
works \citep{Lin2011, Ciesielski2012}, i.e., on short timescales, MHD turbulence is not 
determined by the total pressure.} Therefore, magnetic fluctuations can drive temperature 
variations in the accretion disk, which can lead to UV/optical flux flickering in AGNs. However, the 
magnetic fluctuations at neighboring radii are expected to be nearly independent on timescales 
much less than the viscous timescales (which is about several hundred years at the optical 
emission regions). If so, two consequences are expected: first, the UV/optical inter-band correlations 
should be extremely weak or absent; second, the variability amplitude of UV/optical emission should 
be small since the observed UV/optical emission is an integration of blackbody radiation of many 
radii and the integration largely eliminates the flux variability due to the independence of the fluctuations. 
These predictions are inconsistent with observations (see Section~\ref{sect:intro}). 

\begin{figure}
\plotone{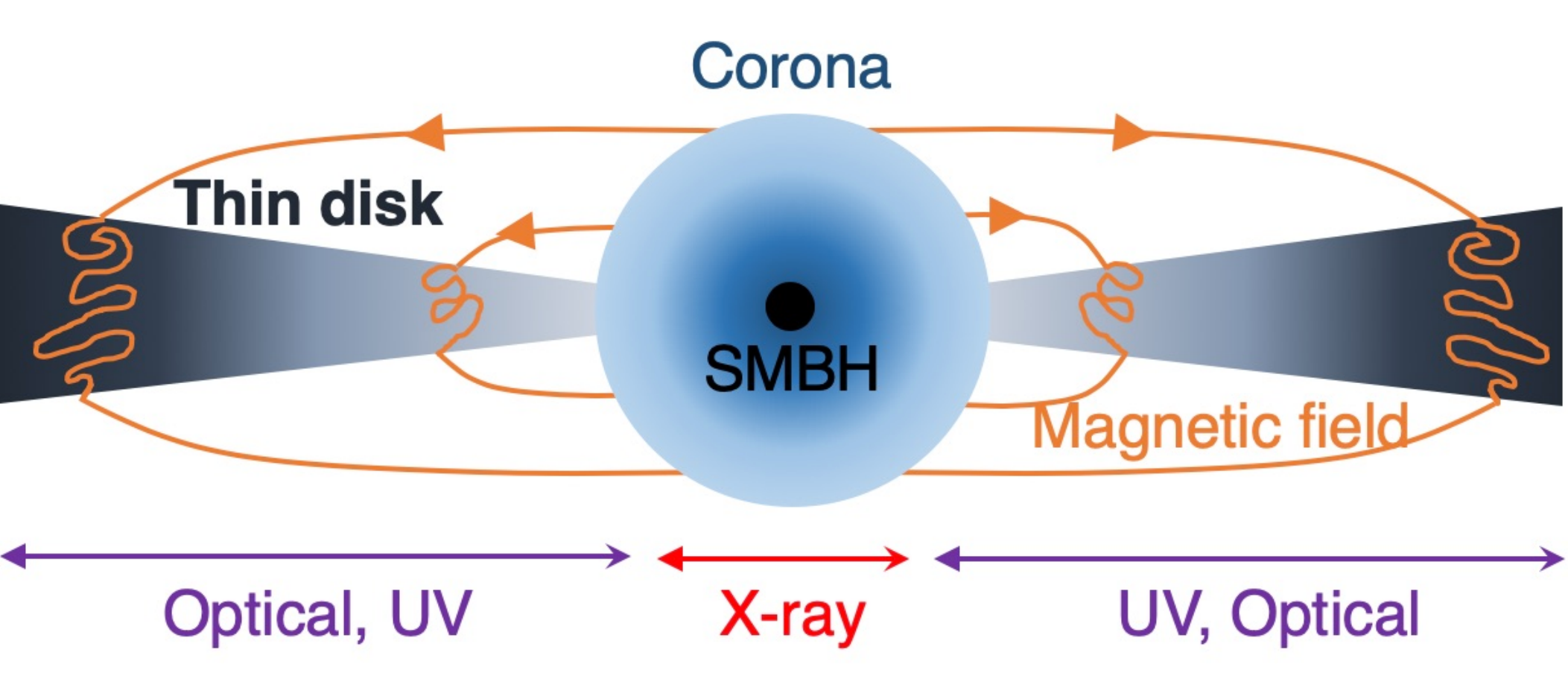}
\caption{Illustration of our CHAR model. The accretion disk 
(gray) and the corona (light blue) are tightly coupled by the magnetic fields (orange curves). 
Note that the disk might extend to the innermost stable radius; these inner regions have 
negligible contribution to the UV/optical variability discussed here. MHD fluctuations/flares 
in the corona can affect disk MHD turbulence and alter the heating rate in the accretion disk. 
As a result, the disk temperature fluctuates in response to the variable heating rate. The 
temperature fluctuations can be determined by solving the equation for thermal-energy 
conservation (i.e., Eq.~\ref{eq:energy}). 
\label{fig:illu}} 
\end{figure} 

To explain the inter-band correlations/time lags and other observational facts of AGN UV/optical 
variability, we propose that the X-ray corona and the underlying accretion disk are tightly coupled 
via the magnetic field (see Figure~\ref{fig:illu}; we defer to Section~\ref{sect:dis1} for a detailed 
discussion of this assumption). As the magnetic field of the corona fluctuates, the magnetic 
fluctuations (with the power of \qill) can also propagate into the accretion disk and induce coherent 
(i.e., the fluctuations at different radii are correlated) disk magnetic turbulence; the disk magnetic 
turbulence dissipates and drives a variable heating rate. As a result, the interior structure of the 
accretion disk should change in response to the variable disk heating rate; the fluctuations of 
disk structure at different radii are also correlated. Hence, when there are chaotic coronal magnetic 
fluctuations/flares, we should expect coherent stochastic variations of the observed UV/optical 
fluxes. 

A mathematical description of the above physical picture is complicated and depends on our complete 
understanding of MHD turbulence. In the absence of such a complete theory of MHD turbulence, 
we have to make a few assumptions to simplify the above physical picture. Without the corona, the time 
average (over the viscous timescale) of the vertically integrated heating rate \qvis \ (which is 
also the dissipation rate of the disk turbulent magnetic power) is 
\begin{equation}
\label{eq:qvis}
\overline{Q_{\mathrm{vis}}^{+}} = \frac{3GM_{\mathrm{BH}}\dot{M}}{4\pi R^3}f_r \\ ,
\end{equation}
where $f_r = 1-\sqrt{3R_{\mathrm{S}}/R}$ and $\dot{M}$ is the absolute accretion rate. In the 
presence of a magnetically coupled compact 
corona, the magnetic fluctuations (with the power per surface area of \qill) from the corona propagate 
into the accretion disk, add to the disk magnetic power, and induce fluctuations of the total magnetic 
power. The heating rate, which is determined by the dissipation rate of the total magnetic power, is 
a summation of \qvis\ and \qill . To specify the total heating rate, we introduce a 
new parameter $k=Q^{+}_{\mathrm{mc}}/Q_{\mathrm{vis}}^{+}$, 
which should be of the order of unity. For simplicity, we may assume that $k$ is constant in $R$; 
it is straightforward to revise our model to consider the case of $k$ as a function of $R$. 

As the total heating rate (which varies in tandem with \qill) changes, the AGN accretion disk should 
be not in vertical hydrostatic equilibrium or thermal equilibrium. The timescale for re-establishing a 
vertical hydrostatic equilibrium is usually much smaller than the timescale for returning to thermal 
equilibrium. If we consider the long-term (i.e., much longer than the thermal timescale) 
variations of the total heating rate, the AGN accretion disk can always adjust its temperature and 
internal energy to re-establish a new (hydrostatic and thermal) equilibrium state, i.e., 
Eq.~\ref{eq:qqq} is valid (but $Q_{\mathrm{X}}$ should be replaced by \qill). Instead, if we consider 
the short-term (i.e., shorter than the thermal timescale) variability of the total heating rate, the 
AGN accretion disk does not have enough time to reach new thermal equilibrium and 
Eq.~\ref{eq:qqq} is inaccurate. 

We assume that an accretion disk can always adjust its vertical structure and scale height to 
respond to the variable \qill\ and \qvis . The temperature fluctuations may be understood 
by solving the vertically integrated thermal-energy conservation law\footnote{As already 
demonstrated by \cite{Lin2012}, Eq.~\ref{eq:energy} can well describe the temperature 
fluctuations due to independent magnetic fluctuations in some MHD shearing-box simulations 
\citep[e.g.,][]{Hirose2009, Jiang2013}.} \citep[see Eq.~(4.58) of][for an accretion disk without 
\qill ]{Kato2008}, 
\begin{equation}
\label{eq:energy}
\frac{\partial E}{\partial t} - (E+\Pi )\frac{\partial \ln \Sigma}{\partial t} + 
\Pi \frac{\partial \ln H}{\partial t} = Q^{+}_{\mathrm{vis}} + Q_{\mathrm{mc}}^{+} - 
Q^{-}_{\mathrm{rad}} \\,
\end{equation}
where $Q^{+}_{\mathrm{vis}}(t)$, $Q_{\mathrm{mc}}^{+}(t-R_X/c_{\mathrm{avf}})$, 
\qrad (t), $E(t)$, $\Pi (t)$, and $\Sigma (t)$ are the vertically integrated internal viscous heating 
rate, the additional variable heating rate due to the presence of the corona, the 
vertically integrated radiative cooling rate, the vertically integrated thermal energy (a summation 
of both gas and radiation), the vertically integrated pressure (a summation of gas pressure 
and radiation pressure), and the surface density, respectively. $Q^{+}_{\mathrm{vis}} + Q_{\mathrm{mc}}^{+}$ 
represent the vertically integrated total heating rate. We assume $Q^{+}_{\mathrm{vis}}$ varies 
in tandem with \qill\ after a magnetic field travel time delay, i.e., $Q^{+}_{\mathrm{vis}}(t)= 
Q_{\mathrm{mc}}^{+}(t-R_X/c_{\mathrm{avf}})/k$. The variability propagation speed along the magnetic 
field is the Alfv\'en velocity, $c_{\mathrm{avf}}$, which depends on the ratio of the magnetic pressure 
to the gas density, and can be close to the speed of light for highly 
magnetized plasma. For simplicity, we assume that the Alfv\'en velocity of the corona-disk coupling 
magnetic field is near to the speed of light since the plasma might be highly magnetized. Note that zero 
advective cooling is assumed, which is a good approximation if the central engine is radiatively 
efficient. 

The solutions of Eq.~\ref{eq:energy} and the expressions for $Q^{-}_{\mathrm{rad}}$, $E$, $\Pi$, and 
$\Sigma$ (which are functions of both time $t$ and radius $R$), which describe the temperature 
fluctuations (and therefore determine UV/optical light curves), depend on the accretion disk model. 
Here we consider the thin disk model of \cite{SSD} with minimum modifications according to MHD 
simulations. We choose the thin-disk model for the following reasons. First, the thin-disk model can 
well fit the SEDs of some AGNs with X-shooter observations \citep{Capellupo2015}. Second, 
the disk-instability model for dwarf-novae and low-mass X-ray binary transients, which is built upon 
the thin-disk model, is widely adopted to adequately explain the outbursts of these systems and 
constrain the viscosity \citep[e.g.,][]{Dubus2001, Lasota2001}. The expressions for \qrad \, $E$, $\Pi$, 
and $\Sigma$ are summarized as follows \cite[for a complete discussion, we refer to Section 
4.4.1 of][]{Kato2008}. The pressure scale height can be determined by 
\begin{equation}
\label{eq:height}
\Omega^2_{\mathrm{K}} H^2 = \frac{\Pi}{\Sigma} \\,
\end{equation}
The vertically integrated pressure, $\Pi$, is 
\begin{equation}
\label{eq:pressure}
\Pi = \Pi_{\mathrm{gas}} + \Pi_{\mathrm{rad}} = \frac{2\kappa_{\mathrm{B}}}{m_{\mathrm{p}}} 
\Sigma T_{\mathrm{c}}+ \frac{2aT^4_{\mathrm{c}}}{3}H \\,
\end{equation}
where $\kappa_{\mathrm{B}}$, $m_{\mathrm{p}}$, $a$, and $T_{\mathrm{c}}$ are the 
Boltzmann constant, the proton mass, the radiation constant, and the inner temperature of 
the accretion disk, respectively. It is often convenient to define 
$\beta \equiv \Pi_{\mathrm{gas}}/\Pi$. The vertically integrated thermal energy, $E$, is simply 
\begin{equation}
\label{eq:int}
E = E_{\mathrm{gas}}+E_{\mathrm{rad}}= \frac{\beta \Pi}{\gamma - 1} + 3(1-\beta)\Pi \\,
\end{equation}
where $\gamma=5/3$ is the ratio of Specific Heat. The vertically integrated radiative cooling 
is 
\begin{equation}
\label{eq:rad}
Q_{\mathrm{rad}}^{-} = \frac{8acT_{\mathrm{c}}^4}{3\tau_{\mathrm{op}}}=2\sigma T_{\mathrm{eff}}^4 \\,
\end{equation}
where $\tau_{\mathrm{op}}$ is optical depth. Optical depth $\tau_{\mathrm{op}}$ is 
\begin{equation}
\label{eq:tau}
\tau_{\mathrm{op}} = \frac{1}{2} (\kappa_{\mathrm{es}}+\kappa_0 \rho T_{\mathrm{c}}^{-3.5}) \Sigma \\,
\end{equation}
where $\kappa_{\mathrm{es}}=0.4\ \mathrm{cm^2g^{-1}}$, and $\kappa_0=6.4\times 
10^{23}\ \mathrm{cm^5 g^{-2} K^{7/2}}$ are the opacity due to electron scattering and free-free 
absorption, respectively. The total opacity is usually dominated by electron scattering at radii 
not larger than $\sim 1000\ R_{\mathrm{S}}$. 

If we focus on timescales smaller than the viscous timescale, \textit{the surface density $\Sigma$ 
can be regarded as constant in time}. Combining Eqs.~\ref{eq:energy}-\ref{eq:int}, we can obtain 
\begin{equation}
\label{eq:Ttime}
C(\beta) \frac{\partial \ln T_{\mathrm{c}}}{\partial t} = \frac{Q_{\mathrm{vis}}^{+}}{\Pi} + 
\frac{Q_{\mathrm{mc}}^{+}}{\Pi} - \frac{Q_{\mathrm{rad}}^{-}}{\Pi} \\,
\end{equation}
where $C(\beta)$ is a function of $\beta$, 
\begin{equation}
\label{eq:cc}
C(\beta) = \Bigg\{ 12(1-\beta) + \frac{\beta}{\gamma-1} + \frac{(4-3\beta)^2}{1+\beta}\Bigg\} \\ .
\end{equation}

In this work, we do not consider any independent non-coherent magnetic fluctuations in the 
accretion disk. Such independent fluctuations are indeed found in accretion-disk MHD 
shearing-box simulations \citep[e.g.,][]{Hirose2009, Jiang2013}. 
The observed AGN luminosity is an integration 
of blackbody radiation of numerous (of the order of $\gtrsim 10^2$) shearing-boxes; the integration 
eliminates the variability of AGN luminosity due to the independent magnetic fluctuations.

\subsection{General Remarks}
\label{sect:result_mk}
Some general features can be inferred from Eq.~\ref{eq:Ttime}. It is convenient to 
define the so-called ``thermal timescale'', i.e.,
\begin{equation}
\label{eq:TH}
\tau_{\mathrm{TH}} \equiv \frac{\overline{E}_{\mathrm{gas}}}{\overline{Q}^{+}_{\mathrm{vis}}} \equiv 
\frac{\overline{\beta}}{\gamma-1} \frac{\overline{\Pi}}{\overline{Q}^{+}_{\mathrm{vis}}} \\,
\end{equation}
where $\overline{E}_{\mathrm{gas}}$, $\overline{Q}^{+}_{\mathrm{vis}}$, $\overline{\beta}$, and 
$\overline{\Pi}$ denote the internal energy of the gas, the viscous heating rate, the ratio of gas 
pressure to total pressure, and the total pressure of a steady solution of Eq.~\ref{eq:Ttime}. 
According to the $\alpha$-prescription of viscosity \citep{SSD} and Eq.~\ref{eq:TH}, 
\begin{equation}
\label{eq:scale}
\tau_{\mathrm{TH}} \sim \frac{1}{\alpha \Omega_{\mathrm{K}}}\propto \alpha^{-1} 
\lambda^2 ((k+1)\dot{M})^{0.5}\propto \alpha^{-1} 
\lambda^2 L_{\mathrm{bol}}^{0.5} \\,
\end{equation}
where $\alpha$, $\lambda$, and $L_{\mathrm{bol}}$ are the dimensionless viscous parameter, 
wavelength, and bolometric luminosity, respectively. This scaling relation can be derived as follows. 
In the steady state, the effective temperature 
profile is (combining Eqs.~\ref{eq:qvis}, \ref{eq:energy}, and \ref{eq:rad} and neglecting 
time variability), $\overline{T}_{\mathrm{eff}}=(3(k+1)GM_{\mathrm{BH}}\dot{M}/(8\pi \sigma k R^3))^{1/4}$. 
For a given wavelength ($\lambda$), its emission region can be estimated by setting $k_{\mathrm{B}} 
\overline{T}_{\mathrm{eff}}=hc/\lambda$, where $h$ and $c$ are the Planck constant and the speed 
of light, respectively. Therefore, it is straightforward to show that $\tau_{\mathrm{TH}}\propto \alpha^{-1} 
\lambda^2 [(k+1)\dot{M}]^{0.5}\propto \alpha^{-1} \lambda^2 L_{\mathrm{bol}}^{0.5}$ since the bolometric 
luminosity $L_{\mathrm{bol}}\propto (k+1)\dot{M}$ (i.e., by integrating the summation of \qvis\ and 
\qill\ over the entire disk). 

In a steady state, we have $\overline{Q}^{+}_{\mathrm{mc}}=\overline{Q}^{+}_{\mathrm{vis}}/k$ and 
$\overline{Q}^{-}_{\mathrm{rad}}=\overline{Q}^{+}_{\mathrm{vis}}+ 
\overline{Q}^{+}_{\mathrm{mc}}$. We can then rewrite Eq.~\ref{eq:Ttime} as 
\begin{equation}
\label{eq:f1}
C(\beta) \frac{\partial \ln T_{\mathrm{c}}}{\partial x} = \frac{\overline{\beta}}{\gamma-1} 
\frac{f_{\mathrm{vis}}^{+}}{f_{\Pi}} + \frac{\overline{\beta}\ k}{\gamma-1}
\frac{f_{\mathrm{mc}}^{+}}{f_{\Pi}} - \frac{\overline{\beta}(k+1)}{\gamma-1} 
\frac{f_{\mathrm{rad}}^{-}}{f_{\Pi}} \\,
\end{equation}
where $x=t/\tau_{\mathrm{TH}}$, $f_{\mathrm{vis}}^{+}= 
Q_{\mathrm{vis}}^{+}/\overline{Q}^{+}_{\mathrm{vis}}$, $f_{\Pi}=\Pi/\overline{\Pi}$, 
$f_{\mathrm{mc}}^{+}=Q_{\mathrm{mc}}^{+}/\overline{Q}^{+}_{\mathrm{mc}}$, and 
$f_{\mathrm{rad}}^{-}=Q_{\mathrm{rad}}^{-}/\overline{Q}^{-}_{\mathrm{rad}}$, respectively. 
Since $Q_{\mathrm{vis}}^{+}$ varies in lockstep with \qill, we expect $f_{\mathrm{vis}}^{+}\equiv 
f^{+}_{\mathrm{mc}}$. Eq.~\ref{eq:f1} can be revised as 
\begin{equation}
\label{eq:final}
f_{\Pi} C(\beta) \frac{\partial \ln T_{\mathrm{c}}}{\partial x} = \frac{\overline{\beta}\ (k+1)}{\gamma-1}
(f_{\mathrm{mc}}^{+}-1) - \frac{\overline{\beta}(k+1)}{\gamma-1} 
(f_{\mathrm{rad}}^{-}-1) \\.
\end{equation}
Note that both $f_{\mathrm{rad}}^{-}$ and $f_{\Pi}$ are functions of $T_{\mathrm{c}}$. For instance, 
let us consider $T_{\mathrm{c}}=\overline{T}_{\mathrm{c}}(1+\delta T)$ with $\delta T \ll 1$. According 
to Eq.~\ref{eq:rad}, $f_{\mathrm{rad}}^{-}=(1+4\delta T)$. For gas-pressure dominated regions (i.e., 
$\beta \simeq 1$), $f_{\Pi}=(1+\delta T)$ (see Eq.~\ref{eq:pressure}); for radiation-pressure dominated 
regions ($\beta \simeq 0$), $f_{\Pi}=(1+8\delta T)$ (see Eqs.~ \ref{eq:height} and \ref{eq:pressure}). 

The second term in the right-hand side of Eq.~\ref{eq:final} can be regarded as a damping term. 
We use $\overline{T}_{\mathrm{c}}$ to represent the inner temperature of a steady solution of 
Eq.~\ref{eq:Ttime}. Suppose that $T_{\mathrm{c}}>\overline{T}_{\mathrm{c}}$, 
then $f_{\mathrm{rad}}^{-}>1$, and the thermal-energy conservation law acts in such a 
way to reduce $T_{\mathrm{c}}$ until $T_{\mathrm{c}}=\overline{T}_{\mathrm{c}}$ (i.e., the 
thermal equilibrium is re-established); the reverse is also true. The characteristic timescale of 
this adjustment is $\sim \tau_{\mathrm{TH}}$. 

The first term in the right-hand side of Eq.~\ref{eq:final} acts as random fluctuations if \qill\ suffers from 
stochastic fluctuations. 
Therefore, Eq.~\ref{eq:final} indicates that the inner temperature $\ln T_{\mathrm{c}}$ is expected to 
vary stochastically with a damping process, which is similar to observed quasar light curves. 
The damping timescale is roughly $\tau_{\mathrm{TH}}$. In fact, since Eq.~\ref{eq:final} is 
(nearly\footnote{This statement is not entirely true because $C({\beta})$ is not scale-free (see 
Eq.~\ref{eq:cc}).}) $\tau_{\mathrm{TH}}$-scale-invariant, we also expect statistical properties of 
$\ln T_{\mathrm{c}}$ fluctuations in AGNs with the same $k$ are similar if the relevant 
timescales are in units of $\tau_{\mathrm{TH}}$ (Figure~\ref{fig:PMS_ALL}; see 
Section~\ref{sect:lbol}).

\section{Results}
\label{sect:result_all}
Throughout Section~\ref{sect:result_all}, the wavelengths and timescales of quasar features are 
always in the rest-frame, unless otherwise specified. 

\subsection{Model Parameters}
\label{sect:model_par}
To understand the temperature fluctuations, we perform numerical calculations. First, we must 
specify the variability behavior of \qill\ (which is presumably produced within $\sim 10\ R_{\mathrm{S}}$). 
Theoretical considerations show that uncorrelated fluctuations at different radii which propagate 
inward result in accretion-power fluctuations in the innermost regions, and the PSD of the 
fluctuations is $\propto 1/f$ \citep[e.g.,][]{Lyubarskii1997, Lin2016}. Three-dimension general 
relativistic MHD simulations \citep{Noble2009} also suggest that the PSD of the corona-energy 
dissipation is $\propto 1/f$. Therefore, we also 
adopt the $1/f$ law as the PSD of \qill . Other PSDs of \qill \ are possible if the fluctuations at different 
parts of the corona are not uncorrelated. Our model can easily be generalized to address other 
PSDs.\footnote{We point out that the UV/optical variability amplitude would be extremely suppressed if 
\qill\ has a flat PSD (i.e., being white noise). In this case, the temperature fluctuations in two adjacent radii are 
independent because they are driven by two different epochs (due to the light-travel time delay) of 
\qill .} The probability density distribution (PDF) of \qill\ is assumed to be log-normal 
\citep[e.g.,][]{Uttley2005}. For illustrative purposes, we fix the fractional variability amplitude of \qill , 
$\delta_{\mathrm{mc}}$, to be $10\%$ on timescales of $10^5$ days in Section~\ref{sect:result_all}; 
our calculations can be easily generalized to consider a larger/smaller variability amplitude. 

\begin{deluxetable}{ccccc}
\tablecaption{Model parameter\label{tbl:tb1}}
\tablehead{\colhead{Case} & \colhead{$M_{\mathrm{BH}}$} & \colhead{$\dot{m}$} 
& \colhead{$L_{\mathrm{bol}}$} \\ 
\colhead{} & \colhead{$M_{\odot}$} & \colhead{} & \colhead{($\mathrm{erg\ s^{-1}}$)}\\ 
(1) & (2) & (3) & (4)} 
\startdata
A &  $10^8$ &  $0.1$ &  $2.5\times 10^{45}$ \\
B &  $10^8$ &  $0.3$ &  $7.6\times 10^{45}$ \\
C &  $10^7$ &  $0.1$ &  $2.5\times 10^{44}$ \\
D &  $5\times 10^7$ &  $0.2$ &  $2.5\times 10^{45}$ \\
\enddata
\tablecomments{Notes. (1) Case. (2) The black hole mass. (3) The dimensionless 
accretion rate $\dot{m}=\dot{M}/\dot{M}_{\mathrm{Edd}}$, where $\dot{M}_{\mathrm{Edd}} 
= 10L_{\mathrm{Edd}}/c^2$. (4) The bolometric luminosity.}
\end{deluxetable}

We must set the initial conditions for Eq.~\ref{eq:final} (or Eq.~\ref{eq:Ttime}). At time $t=0$, the 
initial $T_{\mathrm{c}}$ and $T_{\mathrm{eff}}$ are given by the stationary solution of the standard 
thin accretion disk with additional \qill\ \citep[i.e., by considering the stationary solution of Eq.~\ref{eq:Ttime} 
and the mass, momentum, and angular-momentum conservation laws; see Section~3.2.1 of][]{Kato2008}. 
At this stage, three parameters are introduced, i.e., the dimensionless accretion rate\footnote{The 
dimensionless accretion rate, $\dot{m}$, is the ratio of the accretion rate to the Eddington accretion rate, 
$\dot{m}=0.1c^2\dot{M}/L_{\mathrm{Edd}}$, where $L_{\mathrm{Edd}}$ is the Eddington luminosity. 
Therefore, if the radiative efficiency is $10\%$, $\dot{m}$ also represents the Eddington ratio.} $\dot{m}$, 
$M_{\mathrm{BH}}$, and the viscous parameter $\alpha$. 

The inner and outer boundaries of the accretion disk are fixed to be $10 R_{\mathrm{S}}$ and $1000\ 
R_{\mathrm{S}}$, respectively.\footnote{We note that, when the inner boundary is fixed to be $10 
R_{\mathrm{S}}$, the disk may still extend to the innermost stable radius. The $\lesssim 10 R_{\mathrm{S}}$ 
regions are likely to have negligible contributions to the UV/optical emission we discuss here. Indeed, 
our results remain largely unchanged if we fix the inner boundary to $3R_{\mathrm{S}}$. To model 
the temperature fluctuations in the innermost regions properly, we also must consider general 
relativity effects and complicated comptonization processes. Therefore, we ignore these regions for 
simplicity. } The viscous parameter $\alpha$ is fixed to be $0.1$. Our main conclusions would not 
be significantly changed if we adopted other reasonable values of $\alpha$; the only significant 
change would be the characteristic timescale $\tau_{\mathrm{TH}}$ since this timescale is $\propto 
1/\alpha$. $k$ is assumed to be $1/3$; our results remain unchanged if we consider other values 
of $k$.

\subsection{A Starting Case: $M_{\mathrm{BH}}=10^8\ M_{\odot}$, $\dot{m}=0.1$}
\label{sect:fud}
We start by considering an AGN with $M_{\mathrm{BH}}=10^8\ M_{\odot}$ and $\dot{m}=0.1$ (hereafter case A). 
The bolometric luminosity is $L_{\mathrm{bol}}= 
(1+k)\dot{m} L_{\mathrm{Edd}}=2.52\times 10^{46}\ \mathrm{erg\ s^{-1}}$. 

\begin{figure}
\epsscale{1.2}
\plotone{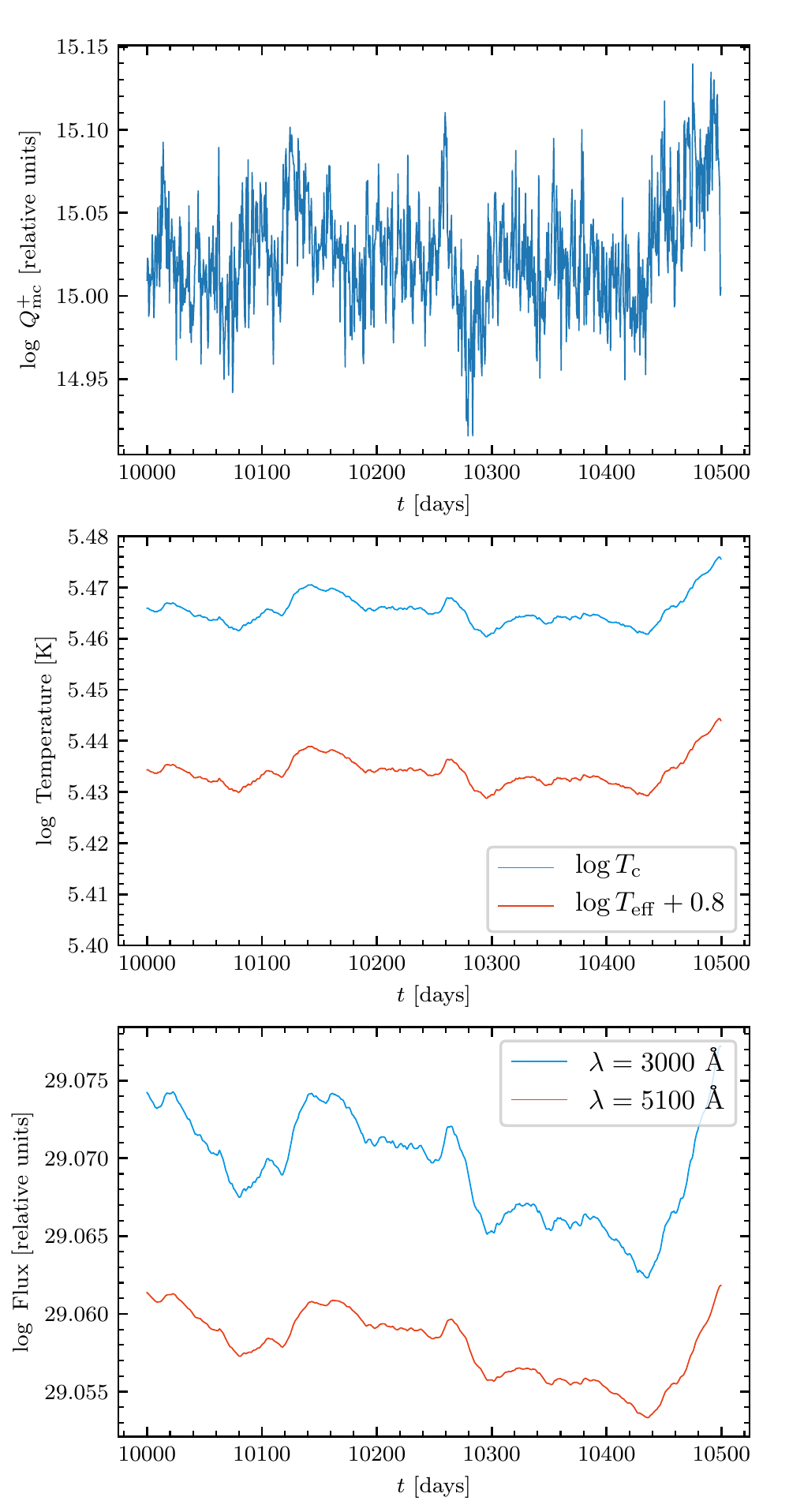}
\caption{Upper: the logarithmic variable heating rate ($\log Q^{+}_{\mathrm{mc}}$) as a function of time. 
Middle: the logarithmic inner temperature ($\log T_{\mathrm{c}}$) and effective temperature ($\log 
T_{\mathrm{eff}}$) at $R_{3000}$ as a function of time. For clarity, we shift $T_{\mathrm{eff}}$ 
upward by $0.8$ dex. 
Lower: the simulated light curves for emission at wavelengths $\lambda=3000\ \mathrm{\AA}$ 
(blue curve) and $\lambda=5100\ \mathrm{\AA}$ (red curve). 
Compared with \qill\ , the fast (i.e., short-term) variability in the UV/optical light curves is 
significantly suppressed. \label{fig:lc_A}}
\end{figure}

We solve Eq.~\ref{eq:Ttime} with a time-step of $0.5$ days to obtain the temporal evolution of 
$T_{\mathrm{c}}$. The total time length of the light curve of $T_{\mathrm{c}}$ is $10^5$ days. 
The effective temperature, $T_{\mathrm{eff}}$, can be derived by considering Eq.~\ref{eq:rad} . 
We then obtain 
the multi-wavelength light curves by integrating the multi-temperature blackbody emission. The light 
curves of \qill\ , $T_{\mathrm{c}}$, $T_{\mathrm{eff}}$, and the $3000\ \mathrm{\AA}$ and $5100\ 
\mathrm{\AA}$ emission are presented in Figure~\ref{fig:lc_A}. For illustrative purposes, we show 
$T_{\mathrm{c}}$ and $T_{\mathrm{eff}}$ at the $3000\ \mathrm{\AA}$ emission characteristic 
radius ($R_{3000}$; i.e., where $k_{\mathrm{B}}T_{\mathrm{eff}}(R_{3000})=h\lambda /c$ with 
$\lambda=3000\ \mathrm{\AA}$). At first glance, the fast (i.e., short-term) variability in the light 
curves of the $3000\ \mathrm{\AA}$ and $5100\ \mathrm{\AA}$ emission is 
significantly suppressed. 

\begin{figure}
\epsscale{1.2}
\plotone{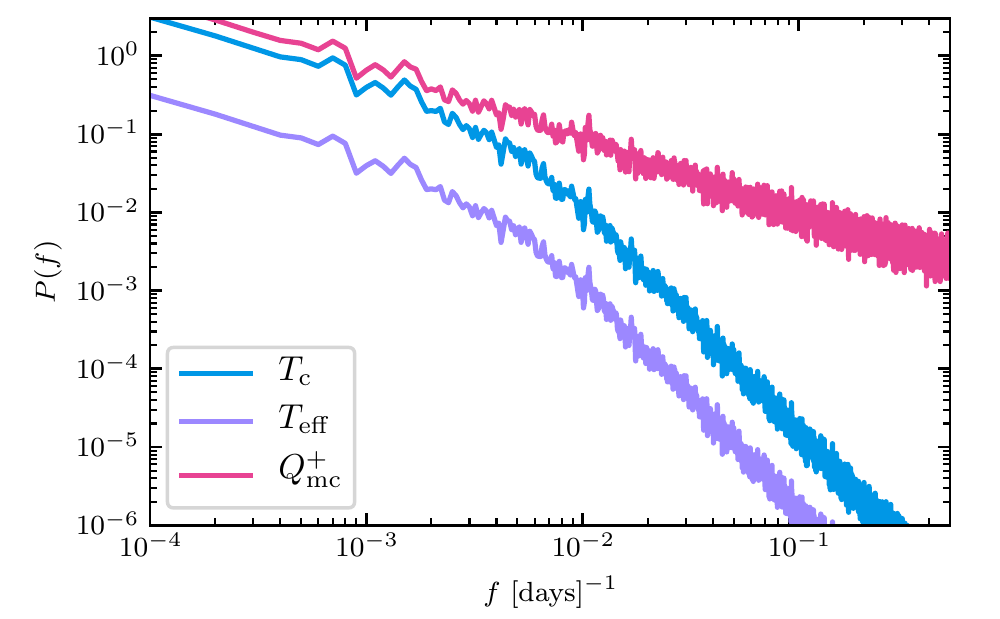}
\caption{The PSDs of $2.5\log Q^{+}_{\mathrm{mc}}$, $2.5\log T_{\mathrm{c}}$ and $2.5 
\log T_{\mathrm{eff}}$. For clarity, we shifted the PSD of $2.5\log T_{\mathrm{c}}$ upward 
by a factor of $10$. 
\label{fig:TPSD}}
\end{figure}

\subsubsection{Statistical Properties of Light Curves}
\label{sect:ssp}
To check the statistical properties of these light curves, we first calculate the PSDs of 
$2.5\log Q^{+}_{\mathrm{mc}}$, $2.5\log T_{\mathrm{c}}$, and $2.5 \log T_{\mathrm{eff}}$. 
We adopt the Welch method \citep{Welch1967} with the light curves broken into ten equal-length segments 
to calculate the PSDs. The results are shown in Figure~\ref{fig:TPSD}. At the low-frequency 
limit, the PSDs of $2.5\log T_{\mathrm{c}}$ and $2.5\log T_{\mathrm{eff}}$ follow that of \qill\ (i.e., 
these PSDs follow the $1/f$ shape) . However, at higher frequencies, the PSDs of $2.5\log T_{\mathrm{c}}$ 
and $2.5 \log T_{\mathrm{eff}}$ are steeper (i.e., having less variability power at higher frequencies) 
than that of \qill\ . 

\begin{figure*}
\epsscale{1.2}
\plotone{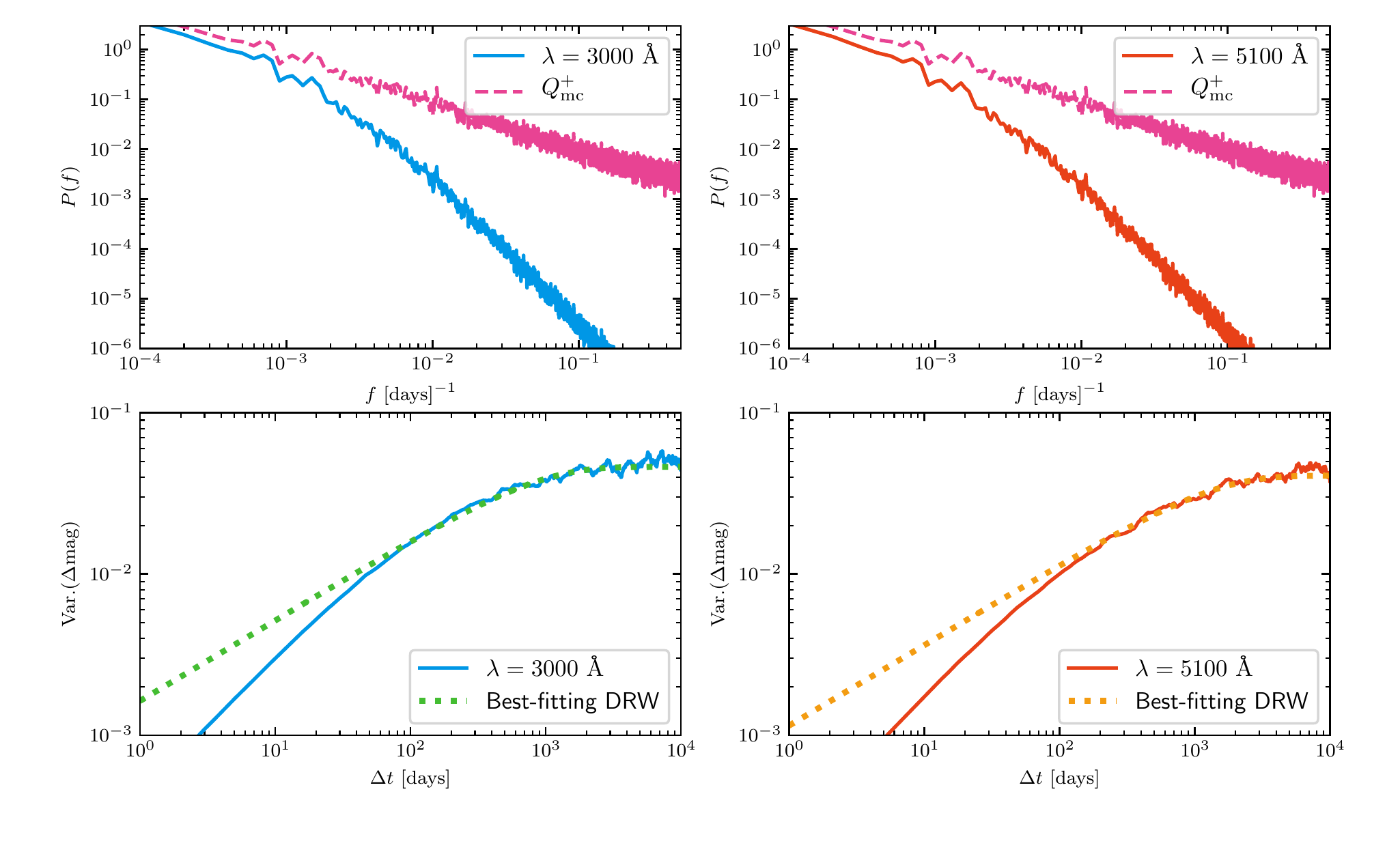}
\caption{Upper-left: The PSDs of the $3000\ \mathrm{\AA}$ emission (blue curve) and $2.5\log Q^{+}_{\mathrm{mc}}$ 
(pink curve). Upper-right: similar to the upper-left one but for the $5100\ \mathrm{\AA}$ emission (red curve). 
Lower-left: The SF of the $3000\ \mathrm{\AA}$ emission (blue curve). The green dotted curve indicates the 
DRW fit to the blue curve. Lower-right: similar to the lower-left one but for the $5100\ \mathrm{\AA}$ emission 
(red curve). The orange dotted curve indicates the DRW fit to the red curve.  On long timescales ($\gtrsim 5000$ 
days), our model is slightly more variable than the predications of the DRW models. However, real observations 
with limited time durations (typically less than $5000$ days or $\sim 14$ years) cannot probe this deviation. 
On very short timescales ($\lesssim 10^2$ days), the SFs of our model are smaller than those of the best-fitting DRW 
fits but are very similar to the $\mathrm{SF}\propto \Delta t^{0.8}$ relation (which is the best-fitting model of the 
best-studied \textit{Kepler} AGN Zw 229-15; see \citealt{Kasliwal2015}). \label{fig:sf_long}}
\end{figure*}

$T_{\mathrm{c}}$ and $T_{\mathrm{eff}}$ are non-observables. Therefore, we now consider the 
statistical properties of light curves of the $3000\ \mathrm{\AA}$ and $5100\ \mathrm{\AA}$ 
emission. We again adopt the Welch method to obtain the PSDs. We also measure the 
structure function (SF) of each light curve; the SF essentially measures the variability amplitude 
as a function of timescale $\Delta t$. It is argued that the SF is more robust than the PSD for 
low/irregular-cadence light curves \citep[for a discussion of SF, please refer to][]{Emmanoulopoulos2010}, 
and it is widely used to quantify AGN UV/optical variability. The SF can be measured by using different 
statistical estimators \citep[e.g.,][]{Sun2015}; each estimator has its own (dis-)advantages (e.g., 
in terms of treatments of observational uncertainties and outliers). When dealing with 
simulated data without any measurement errors, we can use any estimator and choose to 
use the normalized median absolute deviation (NMAD), i.e., 
\begin{equation}
\label{eq:nmad}
\mathrm{SF}(\Delta t) = 1.48 \mathrm{Median}(|\Delta m_{i,j}-\mathrm{Median}(\Delta m_{i,j})|) \\,
\end{equation}
where $\Delta t=|t_i - t_j|$ is the time separation between two observations (with magnitudes 
$m_i$ and $m_j$, respectively) and $\Delta m_{i,j}=m_j - m_i$. 

To calculate the SF, we divide the full light curve of each wavelength into five segments and 
calculate the SF of each segment. For each wavelength, we then average the five SFs to 
obtain our final SF. 

The PSDs and the SFs of the $3000\ \mathrm{\AA}$ and $5100\ \mathrm{\AA}$ light curves are 
shown in Figure~\ref{fig:sf_long}. The SFs show some artificial ``peaks" and ``dips" on timescales 
around $10^4$ days (i.e., the longest timescale that can be probed by our simulated light 
curves); these ``artificial'' features have been identified and discussed by \cite{Emmanoulopoulos2010}. 
Like the PSDs of $T_{\mathrm{c}}$ and $T_{\mathrm{eff}}$, PSDs of the $3000\ \mathrm{\AA}$ and 
$5100\ \mathrm{\AA}$ light curves are steeper than that of \qill\ at high frequencies. 

Motivated by ground-based observations, it has been proposed that the luminosity variations follow 
a DRW process, whose SF is \citep{Kelly2009} 
\begin{equation}
\label{eq:drw_sf}
\mathrm{SF}(\Delta t) = \hat{\sigma}\sqrt{\tau(1-\exp(-\Delta t/\tau))} \\,
\end{equation}
where $\hat{\sigma}$ and $\tau$ are the normalization factor and the damping timescale, respectively. 

\begin{figure}
\epsscale{1.2}
\plotone{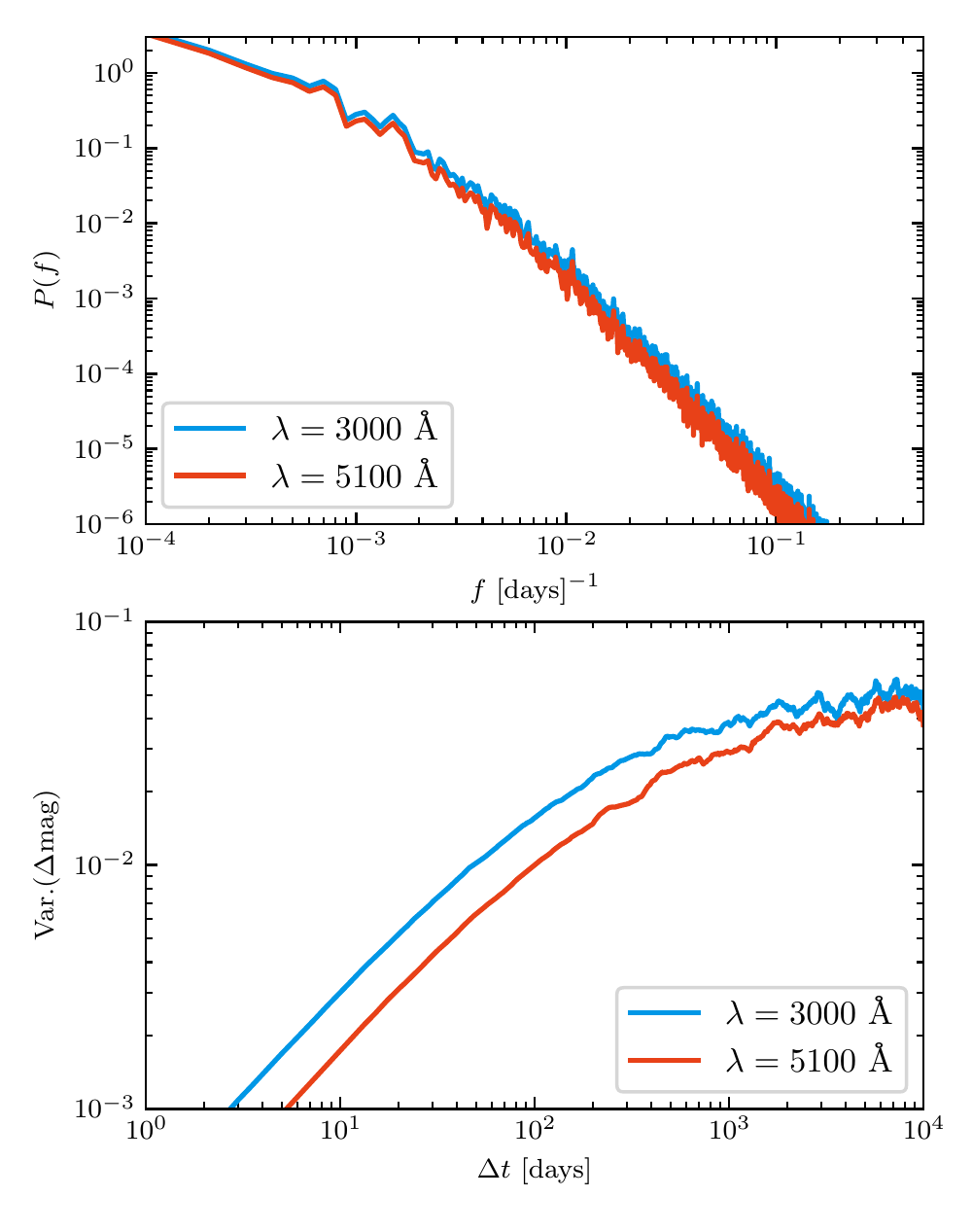}
\caption{Upper: PSDs of the $3000\ \mathrm{\AA}$ (blue curve) and $5100\ \mathrm{\AA}$ 
(red curve) emission. Lower: SFs of the $3000\ \mathrm{\AA}$ (blue curve) and $5100\ 
\mathrm{\AA}$ (red curve) emission. The $3000\ \mathrm{\AA}$ emission is more variable 
(by a factor of $<2$) than the $5100\ \mathrm{\AA}$ emission. 
\label{fig:wave_A}}
\end{figure}

The time duration of observed AGN UV/optical light curves is usually less than $5000$ days 
(i.e., $\sim 14$ years). To compare our results with the DRW model, we fit Eq.~\ref{eq:drw_sf} to 
SFs within timescales smaller than $5000$ days.\footnote{We perform a robust least squares to fit 
the DRW SF to the simulated one; the robust least squares is performed by adopting the function 
``least squares'' in the PYTHON package \textit{scipy} with a ``softl1'' loss function 
and an \textit{fscale} of $0.4$.} The best-fitting DRW SFs are included in the lower panels of 
Figure~\ref{fig:sf_long}. On timescales of $10^2$ up to several thousands of days, the best-fitting DRW model can 
explain the SFs of our light curves. On shorter timescales, the SFs are steeper than the best-fitting 
DRW models, i.e., the DRW models over-predict the short-timescale variability. These results are 
in qualitative agreement with \textit{Kepler} AGN light curves \citep[e.g.,][]{Mushotzky2011}. In fact, 
the SFs of our model on timescales $\lesssim 10^2$ days can be well described by the $\mathrm{SF} 
\propto \Delta t^{0.8}$ relation, which is the best-fitting model of the best-studied \textit{Kepler} AGN 
Zw 229-15 \citep{Kasliwal2015}. The disagreement timescale between the 
DRW process and our model is $\sim 100$ days in Figure~\ref{fig:sf_long}. As mentioned in 
Section~\ref{sect:result_mk}, our model is $\tau_{\mathrm{TH}}$-scale-invariant (see also 
Section~\ref{sect:lbol} and Figure~\ref{fig:PMS_ALL}). If we consider an AGN with $L_{\mathrm{bol}} 
=6.4\times 10^{43}\ \mathrm{erg\ s^{-1}}$ \citep[i.e., the bolometric luminosity of Zw 229-15; 
see][]{Barth2011}, its thermal timescale $\tau_{\mathrm{TH}}$ is expected to be a factor of $6.3$ 
smaller than that of the AGN considered here (see Eq.~\ref{eq:scale}). Therefore, 
the disagreement timescale between the DRW process and our model should be around $100/6.3 
=15.8$ days, which is in agreement with that of Zw 229-15 \citep[see Figure 12 in][]{Kelly2014}. 
Detailed comparisons between our model and the Kepler light curves of Zw 229-15 and several 
other Kepler AGNs are presented in Section~\ref{sect:kepler}.

In addition, on timescales $\gtrsim 5000$ days, the best-fitting DRW models under-predict 
the variability of our light curves. Observationally speaking, there is some indirect evidence that, 
on long timescales, AGNs are more variable than the predication of the DRW model 
\citep{MacLeod2012, Guo2017}.

\subsubsection{Wavelength Dependence}
\label{sect:wave}
\begin{figure}
\epsscale{1.2}
\plotone{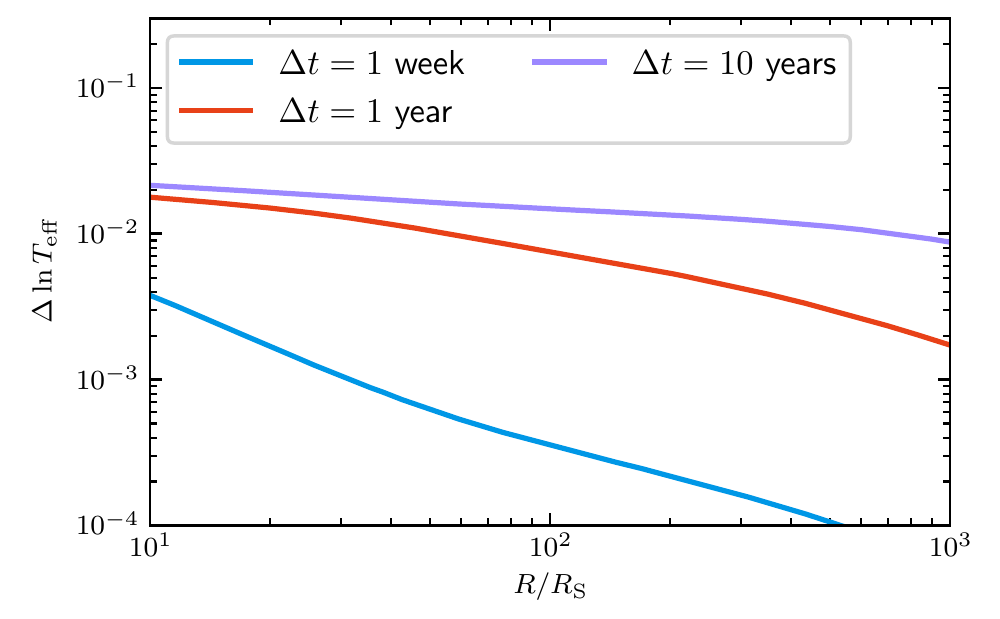}
\caption{The fractional variability amplitude, $\Delta T_{\mathrm{eff}}/T_{\mathrm{eff}}$, as 
a function of radius. The blue, red, and purple solid curves are for timescales of one week, one 
year, and ten years, respectively. The temperature fluctuations are more violent at small radii; 
this tendency is stronger on short timescales. \label{fig:TeffA}}
\end{figure}

We explore the variability amplitude as a function of wavelength and find that the variability 
amplitude declines with increasing wavelength. An example is presented in 
Figure~\ref{fig:wave_A}, which shows the PSDs and SFs of the $3000\ \mathrm{\AA}$ and 
$5100\ \mathrm{\AA}$ emission. Indeed, the shorter/bluer ($3000\ \mathrm{\AA}$) wavelength 
light curve is more variable than the longer/redder ($5100\ \mathrm{\AA}$) wavelength one by 
a factor of $<2$, which is roughly 
consistent with observations \citep{MacLeod2012, Sun2015}; the differences are more evident on 
short timescales. The anti-correlation between the variability amplitude and wavelength 
can be interpreted as follows. The $3000\ \mathrm{\AA}$ emission has a smaller thermal 
timescale than that of the $5100\ \mathrm{\AA}$ emission (see Eq.~\ref{eq:scale}). 
According to Eq.~\ref{eq:final}, for fixed observing timescale $\Delta t$, the variability 
amplitude of $\ln T_{\mathrm{c}}(R_{3000})$ is larger than that of $\ln T_{\mathrm{c}}(R_{5100})$ 
(see Figure~\ref{fig:TeffA}) since $\Delta x=\Delta t/\tau_{\mathrm{TH}}$ of the former is 
larger than the latter. 

\subsubsection{Inter-band Cross Correlation}
\label{sect:ccf}
Inter-band cross correlations and time lags are well expected in our model. We use the cross 
power spectral density \citep[hereafter CPSD; see Section 2.1.2 of][]{Uttley2014} to explore 
inter-band correlations and time lags. We again adopt the Welch method to estimate 
$\mathrm{CPSD}(f)_{1,2}$. $\mathrm{CPSD}(f)_{1,2}$ is usually a complex 
function. The complex modulus reflects the tightness of the correlation between two light 
curves; the complex argument indicates the time lag between two light curves. 

The tightness of the correlation can be obtained by defining coherence, 
\begin{equation}
\label{eq:coh}
\Phi(f)_{12}=\frac{\mathrm{abs}(\mathrm{CPSD}(f)_{1,2})^2}{P(f)_1 P(f)_2} \\, 
\end{equation}
where $\mathrm{abs}(\mathrm{CPSD}(f)_{1,2})$ is the complex modulus of 
$\mathrm{CPSD}(f)_{1,2}$ and $P(f)_1$ and $P(f)_2$ are PSDs of two light curves. 
We find that $\Phi(f)_{12}\cong 0.99$ (i.e., the correlation is tight). 

\begin{figure}
\epsscale{1.2}
\plotone{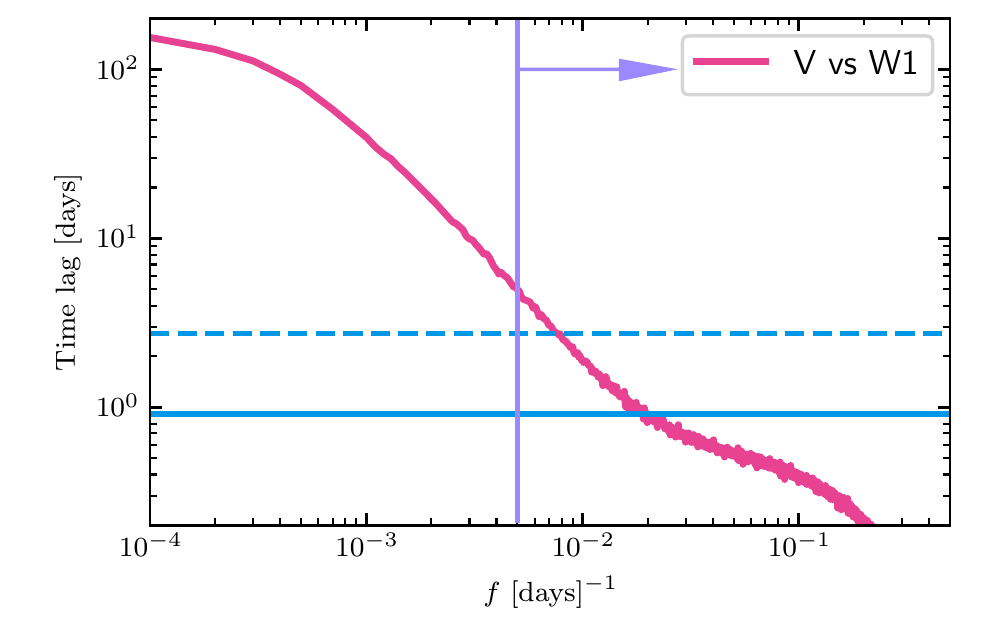}
\caption{The frequency-dependent time lag between the $2700\ \mathrm{\AA}$ (close to the 
central wavelength of the \textit{Swift UVW1} band) and the $5400\ \mathrm{\AA}$ (close to 
the central wavelength of the $V$ band) light curves. The coherence is largely frequency-independent 
and is $\cong0.99$ (i.e., the two light curves are tightly correlated). The blue 
solid line indicates the expected flux-weighted time lag according to the static SSD 
\citep{Fausnaugh2016}. The dashed line corresponds to the time lag if we increase the 
flux-weighted time lag by a factor of three. The durations of many 
high-cadence RM campaigns are $\lesssim 200$ days; therefore, they can only probe 
variations with $f>0.005\ \mathrm{days}^{-1}$ (the purple vertical line with an arrow). 
\label{fig:tlag_A}}
\end{figure}

We then calculate the frequency-dependent time lags from CPSD \citep[see Eq. 10 of][]{Uttley2014},
\begin{equation}
\label{eq:tlag}
\tau(f)_{12}=\frac{\mathrm{arg}(\mathrm{CPSD}(f)_{1,2})}{2\pi f} \\ , 
\end{equation}
where $\mathrm{arg}(\mathrm{CPSD}(f)_{1,2})$ is the argument of the complex variable 
$\mathrm{CPSD}(f)_{1,2}$. 

Unlike the simplest X-ray reprocessing model, the inter-band time lags of our CHAR model 
depend on frequency. 
The frequency-dependent time lags between the $2700\ \mathrm{\AA}$ (close to the central 
wavelength of the \textit{Swift UVW1} band) and $5400\ \mathrm{\AA}$ (close to the 
central wavelength of the $V$ band) light curves are presented in Figure~\ref{fig:tlag_A}. 
To compare with the simplest X-ray reprocessing model, we also show the flux-weighted 
time lag \citep{Fausnaugh2016} of a static SSD with the same $M_{\mathrm{BH}}$ and 
luminosity (hereafter the static SSD time lag\footnote{As demonstrated by \cite{Tie2018}, 
the flux-weighted time lag is $1.5$ times smaller than the expected light-travel time of the 
lamppost X-ray reprocessing model.}). At the high-frequency end (i.e., $f \gtrsim 0.02\ 
\mathrm{days}^{-1}$), the time lag can be less than the static SSD time lag. The physical reasons 
are as follows. The measured time lag is an average of time delays of different radii weighted 
by their surface brightness variations. Unlike the X-ray reprocessing model (which 
assumes constant fractional temperature fluctuations), the fractional temperature fluctuations 
in our model anti-correlate with radius on short timescales (see Figure~\ref{fig:TeffA}) since 
$\tau_{\mathrm{TH}}\propto R^{\frac{3}{2}}$. That is, the inner-disk regions have larger 
fractional temperature fluctuations which can induce larger fractional surface brightness 
variations. Therefore, the weighting factors of inner regions in our model are larger than the 
flux-weighted case, and our model time lag can be smaller than the 
static SSD time lag. On frequencies of $\lesssim 0.01\ \mathrm{day}^{-1}$ (which 
corresponds to the timescales of $\gtrsim 100$ days, i.e., the duration of some high-cadence 
RM campaigns), the time lag approaches the static SSD time lag. Our model time lag can be 
significantly larger than the SSD time lag 
if the frequency is lower than $0.01 \ \mathrm{day}^{-1}$. At extremely low frequencies ($f\sim 
10^{-4}\ \mathrm{day}^{-1}$), the time lag can be $\sim 200$ days, which is roughly the 
difference between the thermal timescale of the $2700\ \mathrm{\AA}$ emission and that 
of the $5400\ \mathrm{\AA}$ emission. Therefore, our model has the potential to explain 
the observed larger-than-expected time lags in some AGNs. 

\begin{figure}
\epsscale{1.2}
\plotone{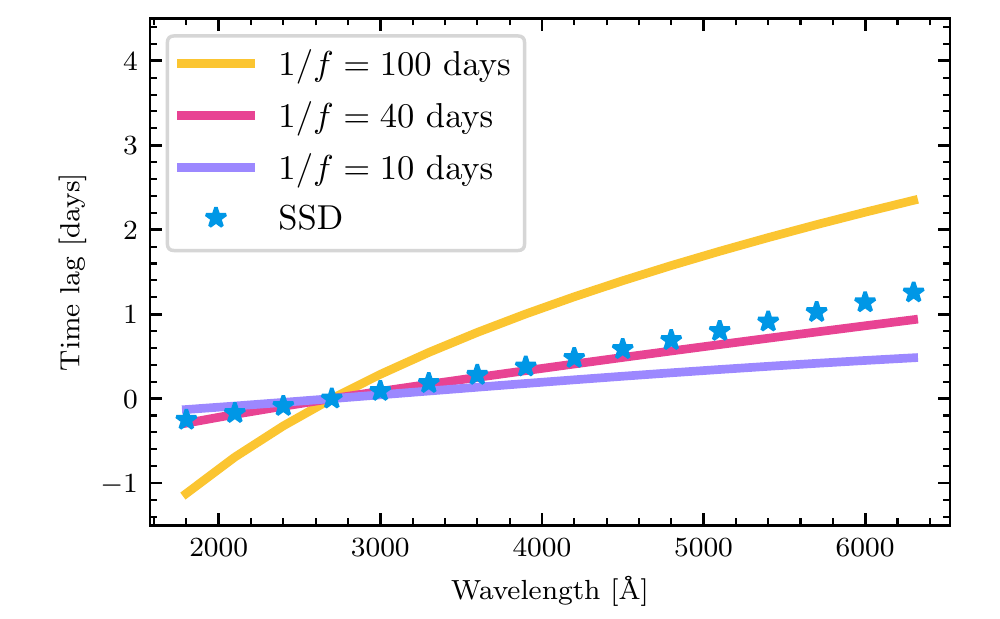}
\caption{The relations between the time lag with respect to the $2700\ \mathrm{\AA}$ (close 
to the central wavelength of the \textit{Swift UVW1} band) emission and wavelength for various 
frequencies. The stars indicate the time lag-wavelength relation for a static SSD. \label{fig:tlag_wave_A}}
\end{figure}

We also calculate the time lag (with respect to the $2700\ \mathrm{\AA}$ emission) as a 
function of wavelength (see Figure~\ref{fig:tlag_wave_A}). Again, it is shown that lower-frequency 
components appear to have larger time lags than those of higher-frequency ones; the slope 
and the normalization of the time lag-wavelength relation also depend on frequency. The 
observed time lag-wavelength relation is an average of various components with different 
frequencies. This average process is complicated and depends on the cadence and duration 
of the RM campaigns. A detailed comparison between our model and the inter-band time lags 
and multi-wavelength SFs of NGC 5548 is presented in Section~\ref{sect:ngc5548}. 

We also point out that UV/optical emission and \qill\ are highly correlated and their time 
lags can also be much larger than the static SSD time lags, especially on long timescales. It 
should be noted that these time lags are not identical to the time lags between X-ray and UV 
emission. This is because there should also be time lags between X-ray and 
\qill . In principle, if the corona can be modeled as an advection dominated accretion flow 
\citep[e.g.,][]{Liu2002, adaf}, the variability of the X-ray emission can also be obtained by solving 
the thermal-energy conservation law of the advection dominated accretion flow. The resulting 
equation is similar to Eq.~\ref{eq:energy} but with an additional advective cooling term on the 
right-hand side of Eq.~\ref{eq:energy}. Meanwhile, unlike the SSD, the 
surface density $\Sigma$ cannot be treated as a constant in time for the advection dominated 
accretion flow. Therefore, the relation between X-ray luminosity and \qill\ can be complicated and 
their variations might not be well coordinated. For instance, an increase in \qill\ may trigger an 
increase in $\Sigma$ or the advective cooling without the necessity 
of invoking an increase in the radiative cooling \qrad\ (i.e., X-ray luminosity). This effect might be 
responsible for the observed weak correlations between X-ray and UV/optical emission 
\citep{Edelson2019}. In the future, we plan to model the relation between X-ray luminosity and \qill\ in 
detail and determine the relation between UV/optical and X-ray emission.

\subsubsection{Microlensing Accretion-Disk Size}
\label{sect:micro}
As mentioned in Section~\ref{sect:intro}, AGN accretion-disk sizes can be measured via microlensing 
observations and the resulting accretion-disk sizes are larger than the flux-weighted radii of the static 
SSD \citep[e.g.,][]{Morgan2010, Cornachione2019}. Our model might be able to resolve this discrepancy. 

Microlensing observations actually measure the half-light radius of the time-variable AGN emission. 
Therefore, we follow \cite{Tie2018} and calculate the half-light radius of the time-variable $3000\ \mathrm{\AA}$ 
emisison as follows. First, we utilize a Taylor expansion to the Planck function and obtain the variation 
of intensity as a function of radius, i.e., 
\begin{equation}
\label{eq:varB}
\Delta B(\lambda, R) = \frac{2hc^2}{\lambda^5} x \frac{\exp(x)}{(\exp(x)-1)^2} 
\frac{\Delta T_{\mathrm{eff}}}{T_{\mathrm{eff}}} \\,
\end{equation}
where $\lambda=3000\ \mathrm{\AA}$, $x=hc/(k_{\mathrm{B}}T_{\mathrm{eff}}\lambda)$ and $B(\lambda)$ 
is the Planck function. Second, for a fixed wavelength, we can calculate the cumulative contribution of 
$\leq R$ regions to the total variability, i.e., 
\begin{equation}
\label{eq:fvar}
f_{\mathrm{\Delta L}}(\lambda, R) = \frac{\int_{10R_{\mathrm{S}}}^{R} \Delta 
B(\lambda, R_0)R_0dR_0}{\int_{10R_{\mathrm{S}}}^{10^3 
R_{\mathrm{S}}} \Delta B(\lambda, R_0)R_0dR_0} \\. 
\end{equation}
Then, the half-light radius, $R_{\mathrm{half}}$, can be calculated by setting $f_{\mathrm{\Delta L}}(\lambda, 
R_{\mathrm{half}}) \equiv 0.5$. 

\begin{figure}
\epsscale{1.2}
\plotone{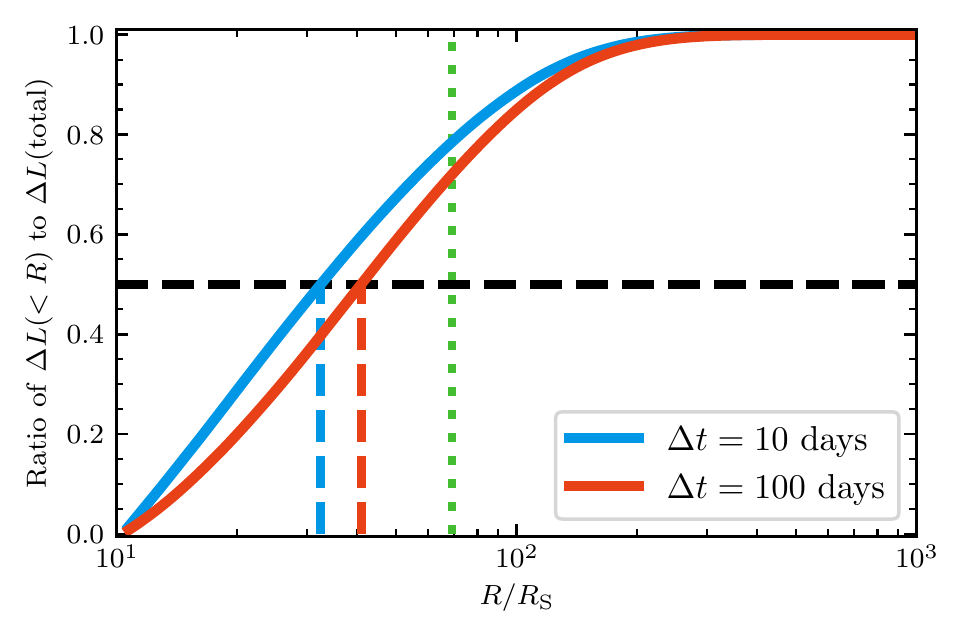}
\caption{The cumulative contribution of $\leq R$ regions to the total variability, $f_{\mathrm{\Delta L}}(\lambda, R)$, as 
a function of $R$. The blue and red curves correspond to timescales of $10$ days and $100$ days, respectively. The 
black dashed line indicates $f_{\mathrm{\Delta L}}\equiv 0.5$. The blue and red dashed lines represent the corresponding 
half-light radii. The green dotted line indicates the half-light radius of the static SSD. In general, the half-light radius of 
our model is smaller than that of the static SSD. \label{fig:rhalf}}
\end{figure}

It is evident that $R_{\mathrm{half}}$ should depend on the relation between $\Delta T_{\mathrm{eff}}/T_{\mathrm{eff}}$ 
and $R$. In the simplest X-ray reprocessing model with a static SSD, it is often assumed that 
$\Delta T_{\mathrm{eff}}/T_{\mathrm{eff}}$ is constant in $R$. In contrast, our model predicts that 
$\Delta T_{\mathrm{eff}}/T_{\mathrm{eff}}$ anti-correlates with $R$ on timescales of $\lesssim 10$ years 
(see Figure~\ref{fig:TeffA}). That is, our half-light radius is smaller than that of the static SSD (see 
Figure~\ref{fig:rhalf}). The flux-weighted radius of our model is similar to that of the static SSD. 
Therefore, our model has the potential to account for AGN microlensing observations 
(i.e., the disk size inferred from the half-light radius of a static SSD is larger than the flux-weighted radius). 
In the future, we plan to convolve our model with gravitational microlensing effects and address the 
microlensing disk size problem in detail. 

\subsubsection{Color Variability}
\label{sect:color}
\begin{figure}
\epsscale{1.2}
\plotone{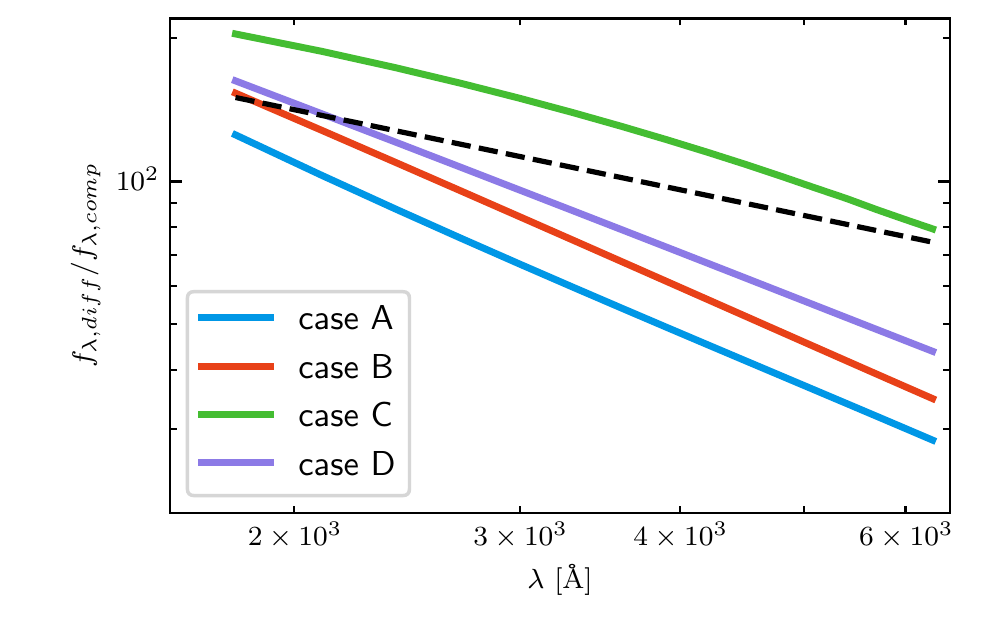}
\caption{The composite differential spectra of our model. The black dashed line corresponds 
to the observed spectrum with an index of $\Gamma_{\lambda}=-0.56$ (\citealt{Ruan2014}). 
The bluer-when-brighter behavior is evident in all cases. The composite differential spectra 
are also similar to the observed one \citep{Ruan2014}. Cases A-D correspond to different 
$M_{\mathrm{BH}}$, $\dot{m}$, and $L_{\mathrm{bol}}$ (see Table~\ref{tbl:tb1}). \label{fig:diff}}
\end{figure}

Observationally, AGN color variations show a bluer-when-brighter tendency \citep{Ruan2014} and this 
tendency is timescale-dependent \citep{SunYH2014}. To check whether our model can predict such a 
timescale-dependent bluer-when-brighter tendency, we also calculate the color variations of our model. 
First, we follow the methodology in Section 5 of \cite{Ruan2014} to obtain $1500$ differential 
spectra. The time separation of two spectra is fixed to be $50$ days. We then scale the $1500$ 
differential spectra to have the same $3000\ \mathrm{\AA}$ emission and use the geometric 
mean to obtain the composite differential spectrum, which is shown in Figure~\ref{fig:diff} (for 
the differences among cases A, B, C, and D, please see Table~\ref{tbl:tb1} and Section~\ref{sect:pms}). 
Our model predicts a bluer-when-brighter power-law spectral variability that is quite similar to 
the observed one. 

\begin{figure}
\epsscale{1.2}
\plotone{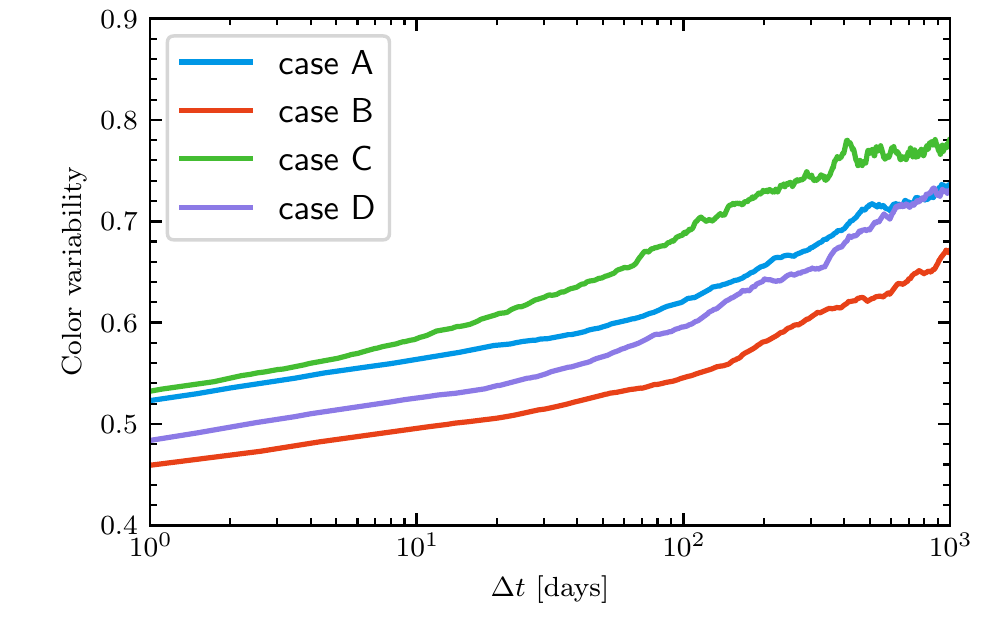}
\caption{AGN time-dependent color variability of our model. The bluer-when-brighter behavior is more 
evident on short timescales than on long timescales. \label{fig:vcolor}}
\end{figure}

We also calculate the timescale-dependent color variability ($S(\Delta t)$), which measures 
the ratio of the variations of the shorter-wavelength emission to those of the longer one, by following 
the methodology in Section 4 of \cite{Zhu2018}. The results of the color variability between the 
$3000\ \mathrm{\AA}$ and $5100\ \mathrm{\AA}$ emission are shown in Figure~\ref{fig:vcolor}. 
Indeed, our model also predicts the timescale-dependent bluer-when-brighter behavior, i.e., the 
bluer-when-brighter behavior is also less prominent on long timescales. 

\subsection{Parameter Dependence}
\label{sect:pms}
According to our model, AGN UV/optical variability depends upon AGN physical parameters, 
namely, $M_{\mathrm{BH}}$ and $\dot{m}$. Therefore, we solve Eq.~\ref{eq:Ttime} 
for four cases (see Table~\ref{tbl:tb1}). 

\subsubsection{Dimensionless Accretion-Rate Dependence}
\label{sect:mdot}
To explore the relation between AGN UV/optical variability and $\dot{m}$, we compare case A 
with two cases (i.e., cases B and D). Cases A and B have the same $M_{\mathrm{BH}}$ but 
different $\dot{m}$ and $L_{\mathrm{bol}}$. On the other hand, cases A and D share the same 
$\dot{M}$ and $L_{\mathrm{bol}}$ but different $\dot{m}$ and $M_{\mathrm{BH}}$. 

\begin{figure}
\epsscale{1.2}
\plotone{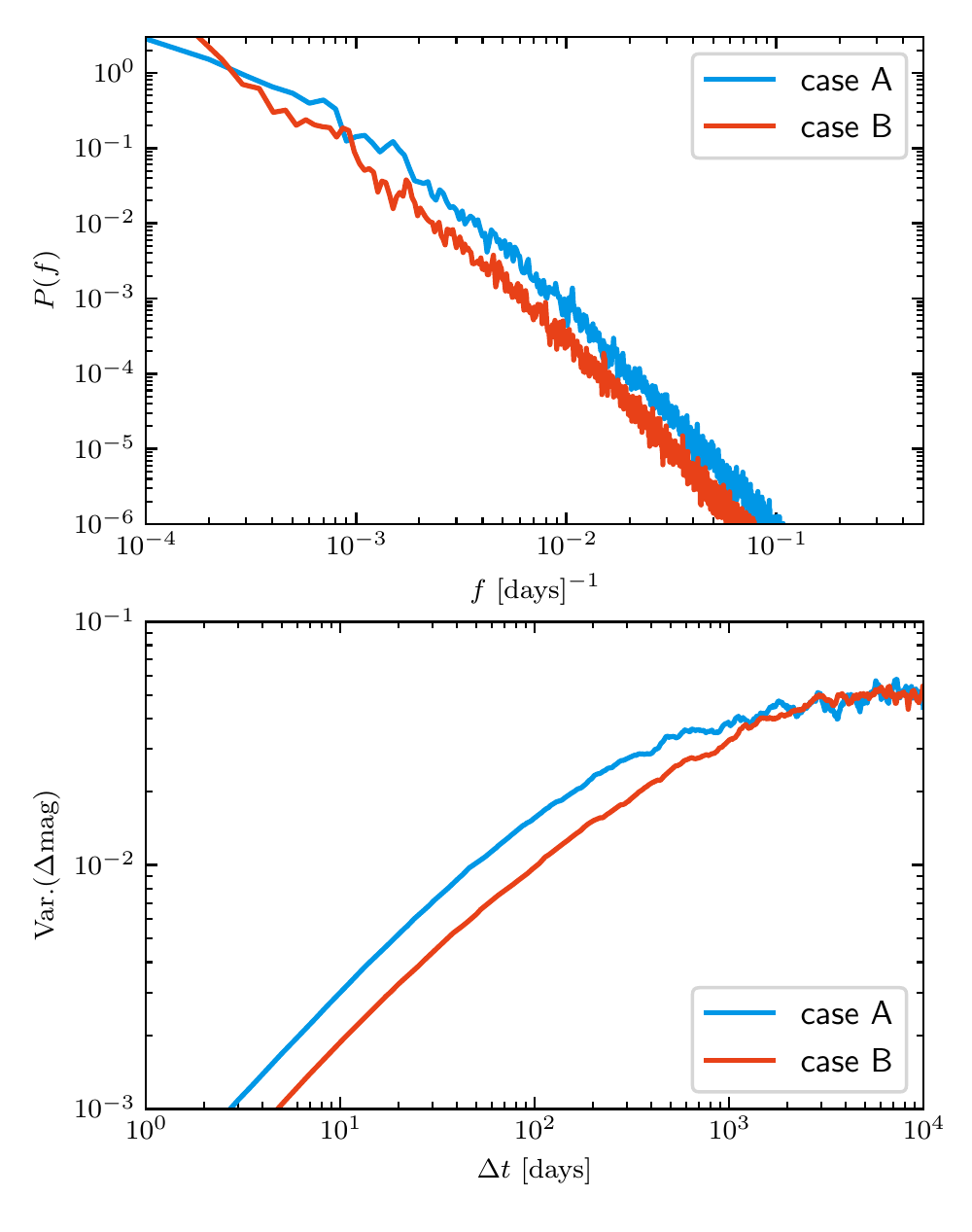}
\caption{Upper: PSDs of the $3000\ \mathrm{\AA}$ emission for cases A and B. Lower: SFs of 
the $3000\ \mathrm{\AA}$ emission for cases A and B. For fixed $M_{\mathrm{BH}}$, the 
variability amplitudes on timescales $\lesssim 10^3$ days decrease with 
increasing $\dot{m}$ or luminosity. This anti-correlation seems to be 
consistent with the empirical relation of \cite{MacLeod2010} for $\hat{\sigma}$ (see text in 
Section~\ref{sect:mdot}). 
\label{fig:PMS_AB}}
\end{figure}

\begin{figure}
\epsscale{1.2}
\plotone{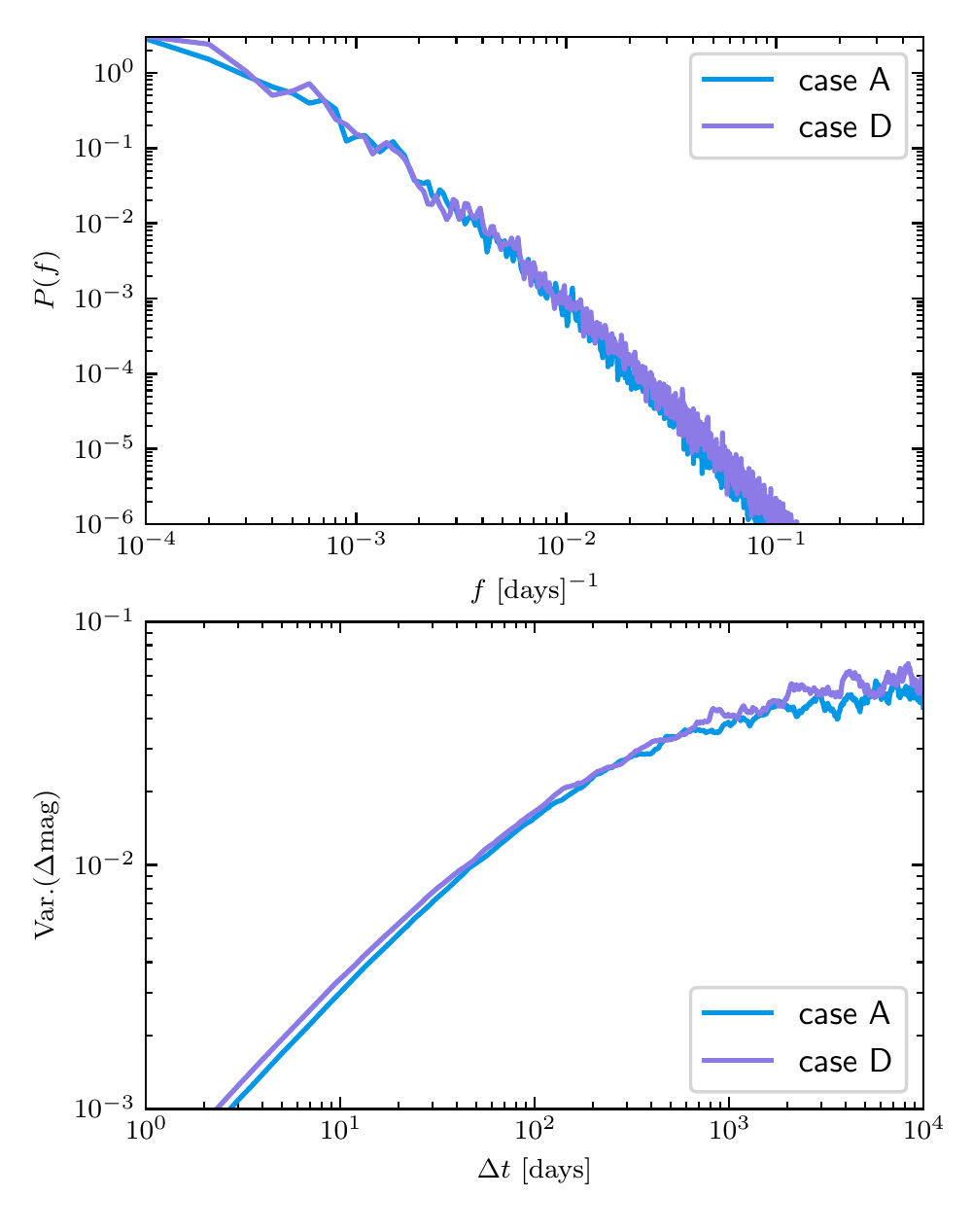}
\caption{Upper: PSDs of the $3000\ \mathrm{\AA}$ emission for cases A and D. Lower: SFs of 
the $3000\ \mathrm{\AA}$ emission for cases A and D. For fixed $L_{\mathrm{bol}}$, the 
variability amplitudes are insensitive to $\dot{m}$ or $M_{\mathrm{BH}}$, which 
is roughly consistent with the empirical relation of \cite{MacLeod2010}. \label{fig:PMS_AD}}
\end{figure}

For illustrative purposes, we focus on the statistical properties of the light curves of the $3000\ 
\mathrm{\AA}$ emission. A comparison between the PSD and SF of case A and those of cases 
B and D is presented in Figures~\ref{fig:PMS_AB} and \ref{fig:PMS_AD}. 

Empirical relations between the observed SFs and AGN physical properties (e.g., 
luminosity and $M_{\mathrm{BH}}$; for a summary, refer to the Introduction section) have 
been obtained. One popular empirical relation was reported by \cite{MacLeod2010}. It should be noted 
that such empirical relations might suffer from various statistical biases \citep[for a detailed discussion, 
see, e.g.,][]{Kozlowski2017}. However, the empirical relation of \cite{MacLeod2010} for the 
DRW parameter $\hat{\sigma}$ (which determines the SF of a DRW model at timescales that are 
much smaller than the damping timescale) is likely to be reliable \citep[refer to Figure 1 
of][]{Kozlowski2017}. Therefore, our model predictions 
are compared with this empirical relation, i.e., $\hat{\sigma} \propto L^{-0.29}M_{\mathrm{BH}}^{0.075} 
\propto \dot{m}^{-0.29}M_{\mathrm{BH}}^{-0.215}$ \cite[see Section~5.2 of][]{MacLeod2010}.\footnote{A 
full comparison between our model and the popular empirical relations, which considers statistical 
biases due to, e.g., irregular and sparse sampling, will be investigated in future works.}

On timescales of $\lesssim 10^3$ days, the $3000\ \mathrm{\AA}$ emission of case A is more 
variable than that of case B by a factor of about $1.5$. According to the empirical relation 
of \cite{MacLeod2010} for $\hat{\sigma}$, the SF of case A is expected to be larger than that of 
case B by a factor of $3^{0.3}\simeq1.4$. That is, for fixed $M_{\mathrm{BH}}$ (and \qill), AGN 
UV/optical variability decreases with increasing $\dot{m}$ or $L_{\mathrm{bol}}$, and 
this anti-correlation is roughly consistent with the empirical relation of \cite{MacLeod2010}. Cases 
A and D have similar SFs, i.e., for fixed $L_{\mathrm{bol}}$ (or $\dot{M}$), AGN UV/optical variability 
is insensitive to $\dot{m}$ or $M_{\mathrm{BH}}$, which is again consistent with the 
empirical relation of \cite{MacLeod2010}. 

As for color variability, the differential spectra of cases B and D share a similar shape 
with that of case A (Figure~\ref{fig:diff}). However, cases B and D show 
more evident bluer-when-brighter behaviors than case A (Figure~\ref{fig:vcolor}). That is, the 
timescale-dependent color variability (bluer-when-brighter) correlates with $\dot{m}$. 

\subsubsection{Black-Hole Mass Dependence}
\label{sect:mbh}
To explore the relation between AGN UV/optical variability and $M_{\mathrm{BH}}$, 
we compare case A with two cases (i.e., cases C and D). Cases A and C have the same 
$\dot{m}$ but different $M_{\mathrm{BH}}$ and $L_{\mathrm{bol}}$. On the other 
hand, cases A and D share the same $L_{\mathrm{bol}}$ but different $\dot{m}$ 
and $M_{\mathrm{BH}}$. 

\begin{figure}
\epsscale{1.2}
\plotone{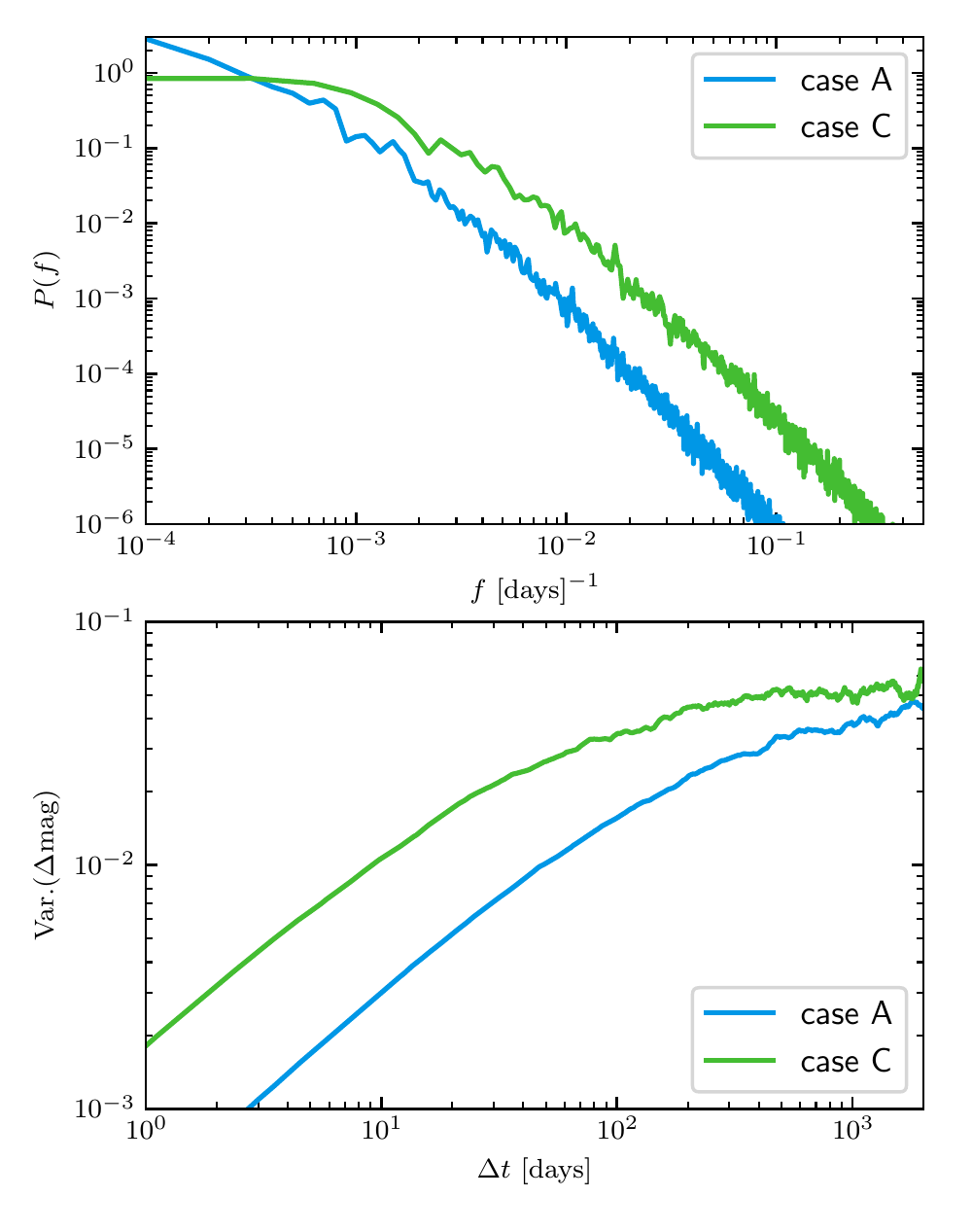}
\caption{Upper: PSDs of the $3000\ \mathrm{\AA}$ emission for cases A and C. Lower: SFs of 
the $3000\ \mathrm{\AA}$ emission for cases A and C. For fixed $\dot{m}$, the variability 
amplitudes on timescales $\lesssim 10^3$ days decrease with increasing $M_{\mathrm{BH}}$ 
or luminosity. This anti-correlation seems to be roughly consistent 
with the empirical relation of \cite{MacLeod2010} for $\hat{\sigma}$ (see texts in 
Section~\ref{sect:mbh}). 
\label{fig:PMS_AC}}
\end{figure}

Compared with case A, case C predicts larger variability amplitudes (by a factor 
of about $2.2$) of the $3000\ \mathrm{\AA}$ emission on timescales of $\lesssim 10^3$ days 
(Figure~\ref{fig:PMS_AC}). Therefore, if we control $\dot{m}$ (and \qill), AGN UV/optical variability 
and $M_{\mathrm{BH}}$ are anti-correlated. According to the empirical relation of 
\cite{MacLeod2010} for $\hat{\sigma}$, the SF of case C is expected to be larger than that of 
case A by a factor of $10^{0.215}=1.64$, which is smaller than our model prediction by a factor 
of $2.2/1.64=1.34$. This small discrepancy might be understood as follows. The sample of 
\cite{MacLeod2010} consists of luminous AGNs while the AGN of case C has a much lower 
bolometric luminosity ($2.5\times 10^{44}\ \mathrm{erg\ s^{-1}}$). There is some evidence to 
suggest that the empirical relation of \cite{MacLeod2010} under-predicts the short-term variability 
amplitudes (by a factor of $\sim 1.3$) of AGNs with relatively low luminosities \citep[see Section 6.1 
of][]{Sun2015}. 

Instead, if we fix $L_{\mathrm{bol}}$ (i.e., case A vs case D; 
Figure~\ref{fig:PMS_AD}) and \qill, there is no strong correlation between AGN 
UV/optical variability and $M_{\mathrm{BH}}$ which is again consistent with the 
empirical relation of \cite{MacLeod2010}. 

As for color variability, the differential spectra of cases C and D share roughly the same shape 
with that of case A (Figure~\ref{fig:diff}). However, case C (D) shows weaker (stronger) 
timescale-dependent color variability (bluer-when-brighter) than case A. Therefore, the dependence 
of AGN timescale-dependent color variability upon $M_{\mathrm{BH}}$ is complicated. 

\begin{figure}
\epsscale{1.2}
\plotone{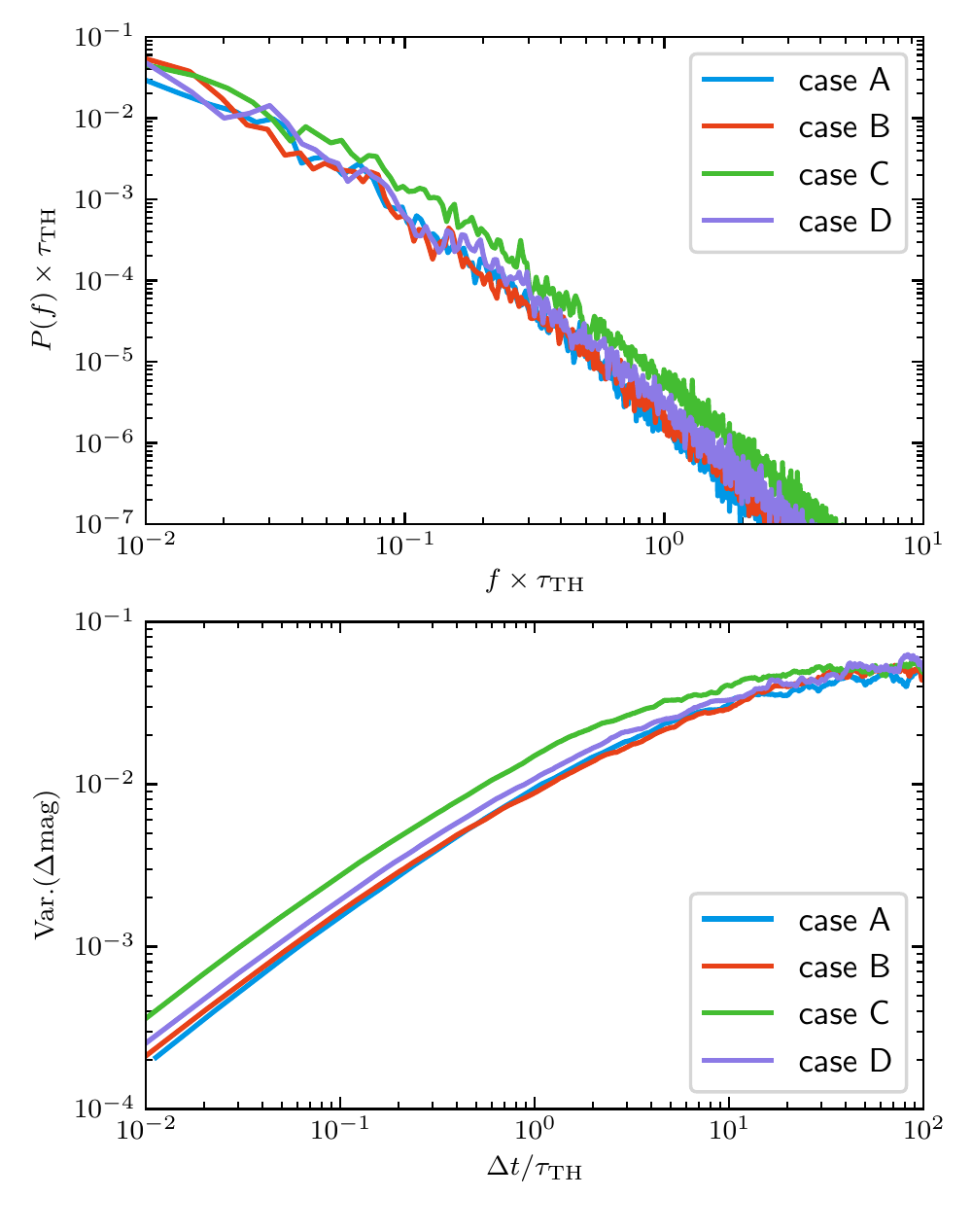}
\caption{Upper: PSDs of the $3000\ \mathrm{\AA}$ emission for cases A, B, C and D. Lower: 
SFs of the $3000\ \mathrm{\AA}$ emission for cases A, B, C and D. When $\Delta 
t$ (the frequency $f$) is expressed in units of the thermal timescale $\tau_{\mathrm{TH}}$ (the 
thermal frequency $1/\tau_{\mathrm{TH}}$), the PSDs and SFs are nearly the same. The 
PSDs and SFs depend weakly on $M_{\mathrm{BH}}$ (see case C). 
\label{fig:PMS_ALL}}
\end{figure}

 \subsubsection{Luminosity Dependence}
 \label{sect:lbol}
From Sections~\ref{sect:mdot} and \ref{sect:mbh}, we can conclude that the PSD and SF of 
the $3000\ \mathrm{\AA}$ emission depend mostly on AGN luminosity. This tendency can 
be understood as follows. Eq.~\ref{eq:Ttime} is roughly scale-invariant if timescale is in units of the 
thermal timescale $\tau_{\mathrm{TH}}$ (see Eq.~\ref{eq:final}) at the $3000\ \mathrm{\AA}$ 
emission characteristic  radius ($R_{3000}$);  as mentioned in Section~\ref{sect:result_mk}, 
$\tau_{\mathrm{TH}}(\lambda)\propto \alpha^{-1}\lambda^2 L_{\mathrm{bol}}^{0.5}$. 
Indeed, if we express timescale in units of $\tau_{\mathrm{TH}}$ days, SFs and PSDs 
of cases A, B and C are quite similar to those of case D (Figure~\ref{fig:PMS_ALL}).\footnote{There 
is a weak dependence upon $M_{\mathrm{BH}}$ (see case C). This is because $C(\beta)$ 
in Eq.~\ref{eq:final} relies on $M_{\mathrm{BH}}$.} This 
feature might be responsible for the observed tight correlation between the short-term 
variability amplitude and AGN luminosity \citep[e.g.,][]{MacLeod2010, Sun2018c}. 

It is evident that AGN timescale-dependent color variability (bluer-when-brighter) correlates with 
$L_{\mathrm{bol}}$ or $\dot{M}$ (i.e., by comparing case A with cases B and C; see 
Figures~\ref{fig:diff} and \ref{fig:vcolor}).

\subsubsection{Inter-band Time Lags and AGN Parameters}
\begin{figure}
\epsscale{1.2}
\plotone{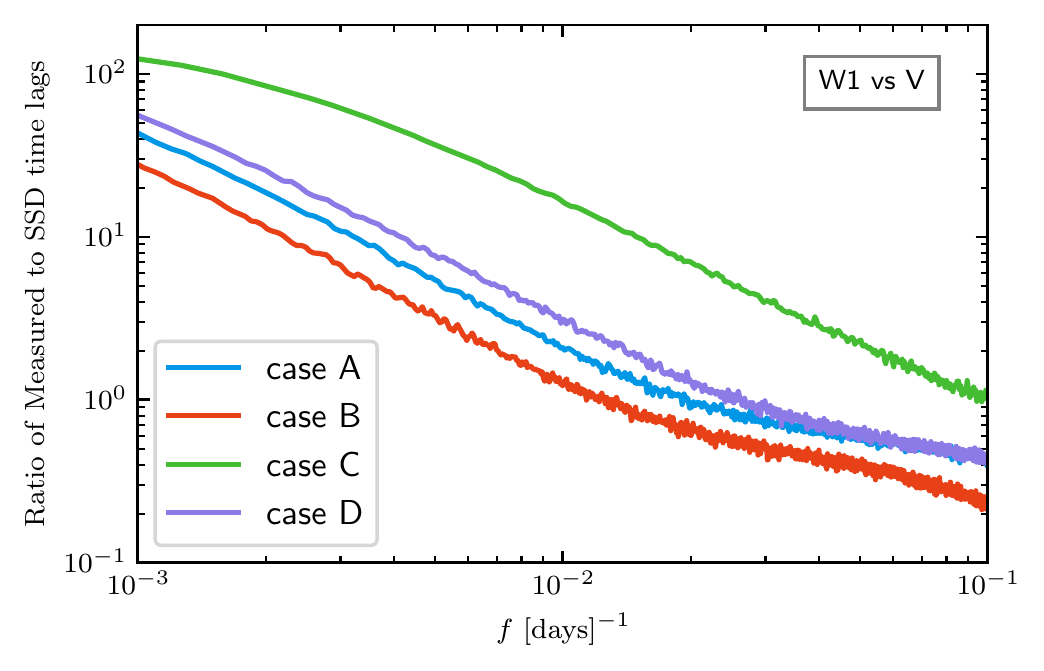}
\caption{The ratios of the frequency-dependent time lags of the UVW1 emission with respect to 
V-band emission to the static SSD time lags for cases A, B, C, and D. Positive lags indicate that 
V band lags UVW1 band. For fixed frequency, less luminous sources tend to have larger ratios 
of measured-to-expected time lags than more luminous ones. 
\label{fig:tlag_ALL}}
\end{figure}

To complete our study, we show the ratios of the time lags between the UVW1-band and the V-band 
emission to the expectations of a static SSD (with the same $M_{\mathrm{BH}}$ and $\dot{m}$) as 
a function of frequency for cases A, B, C and D. The results are presented in Figure~\ref{fig:tlag_ALL}. 
For fixed observing duration $T_{\mathrm{dur}}=1/f_{\mathrm{dur}}$, the ratio of measured-to-expected 
time lag of a less luminous AGN is larger than that of a more luminous one (see Figure~\ref{fig:tlag_ALL}); 
this can explain observational fact \# 6, i.e., the larger-than-expected time lags are observed for 
several local Seyfert 1 AGNs but not in some more luminous and distant AGNs. 

More luminous AGNs tend to have larger thermal timescales (for fixed wavelength and $\alpha$). As a 
result, the anti-correlation between the fractional variability of the effective temperature and radius is 
more evident for more luminous AGNs than for less luminous ones. That is, the microlensing 
accretion-disk size (i.e., the half-light radius of the time-variable emission; see Section~\ref{sect:micro}) 
is always over-estimated if the static SSD model is adopted when studying AGN microlensing observations.

\section{Confronting Our CHAR Model with Real Observations}
\label{sect:test}
We apply our CHAR model to explain two different sets of real observations. 
First, we consider the long-duration ($\sim 3$ years), high-cadence ($\sim 30$ min) Kepler light curves 
of three AGNs (namely Zw 229-15, kplr 12158940 and kplr 2694186) which have reliable black-hole mass 
measurements \citep{Barth2011,Smith2018}. Second, we focus on the multi-wavelength light curves of 
NGC 5548. Throughout Section~\ref{sect:test}, the wavelengths and timescales of quasar features are 
always in the observed-frame, unless otherwise specified. 
\subsection{Kepler Observations}
\label{sect:kepler}
The \textit{Kepler} space telescope \citep{Borucki2010} provided extremely high-cadence ($\sim 30$ min) and 
long-term ($\sim 3$ years) optical light curves for about two dozen AGNs \citep{Smith2018}. Previous works 
using such \textit{Kepler} light curves have revealed that AGN short-term (i.e., $\lesssim 10$ days) variability is 
inconsistent with the DRW model \citep{Mushotzky2011,Kasliwal2015} although this 
model has been proven to be very useful in describing more limited ground-based data 
\citep[e.g.,][]{Kelly2009, MacLeod2010}. 

We select \textit{Kepler} AGNs with $M_{\mathrm{BH}}$ estimates (via the reverberation-mapping or the 
single-epoch virial black-hole mass estimators) from \cite{Barth2011} and Table 1 of \cite{Smith2018}. The 
$\sim 3$-year \textit{Kepler} light curves were broken into multiple segments due to instrumental effects. To ensure 
that the \textit{Kepler} data can efficiently probe both short-term 
and long-term variability, we reject AGNs with data from less than $10$ segments. At this stage, five AGNs 
are selected, namely Zw 229-15, kplr 2694186, kplr 12158940, kplr 12208602, and kplr 9650712. Among 
them, kplr 12208602 is a radio-loud AGN, i.e., non-disk jet emission might be important; and kplr 9650712 
might show a quasi-periodic oscillation signal \citep{Smith2018b} which is beyond the scope of this work. 
Therefore, we do not consider these two AGNs, either. Our final sample consists of three 
sources, Zw 229-15, kplr 2694186 and kplr 12158940. 

\begin{deluxetable*}{ccccccc}
\tablecaption{Quality-of-fit assessment for the three AGNs \label{tbl:tb2}}
\tablewidth{0pt}
\tablehead{\colhead{Name} & \colhead{$\log M_{\mathrm{BH}}$} & \colhead{$\dot{m}$} & \colhead{$\alpha$} & 
\colhead{$\sigma_{\mathrm{mc}}$} 
& \colhead{$\chi^2/D.O.F.$} & \colhead{$P(\chi^2_{\mathrm{mc}}>\chi^2_{\mathrm{obs}})$} \\
\colhead{ } & \colhead{$M_{\odot}$} & \colhead{ } & \colhead{ } & \colhead{ } 
& \colhead{ } & \colhead{}}
\decimalcolnumbers
\startdata
{Zw 229-15}     & {7.00} & {0.034}  & 0.20  &  0.065   & 0.53 & 0.67             \\ 
{       }  & {      } &  {     } & 0.01 & 0.310   & 3.00  & 0.05          \\ 
\hline
{kplr 12158940} & {8.04} & {0.002}  & 0.20  & 0.084   & 3.35  & 0.02           \\
{                 }    & {   } & {   }  &0.01  & 0.400    & 0.23   & 0.92          \\ 
\hline
{kplr 2694186}  & {7.66} & {0.043} & 0.20   & 0.074   &  2.71 & 0.03         \\ 
{                    }     & {   } &  {   }  & 0.01  &  0.640   & 1.77  & 0.15          \\ 
\enddata
\tablecomments{Notes. (1) Object name. (2) The black hole mass (for Zw 229-15, see \cite{Barth2011}; 
for others, see \cite{Smith2018}). (3) The dimensionless accretion rate $\dot{m}=\dot{M}/\dot{M}_{\mathrm{Edd}}$, 
where $\dot{M}_{\mathrm{Edd}}=10L_{\mathrm{Edd}}/c^2$. (4) The dimensionless viscosity parameter. (5) The 
variability amplitude of $Q_{\mathrm{heat}}^{+}$. (6) The ratio of $\chi^2$ to degree of freedom (D.O.F., which 
is $4\times 10^4$). (7) $P(\chi^2_{\mathrm{mc}}>x)$ is the survival function of the distribution of 
$\chi^2_{\mathrm{mc}}$.}
\end{deluxetable*}

\begin{figure}
\plotone{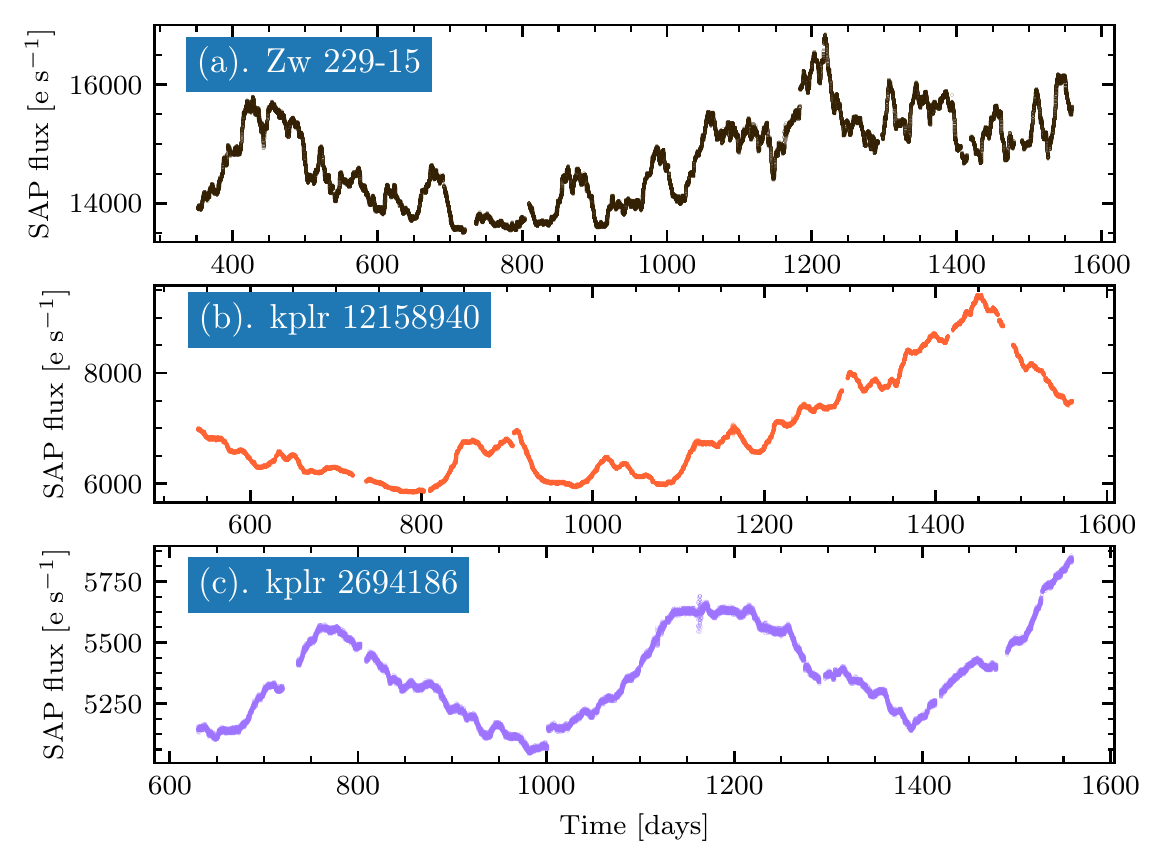}
\caption{The \textit{Kepler} light curves for the three \textit{Kepler} AGNs with high-quality data (with 
a timespan of $\sim 3$ years and a cadence of $\sim 30$ min)  and $M_{\mathrm{BH}}$ estimates. The 
typical fractional simple aperture photometry (SAP) flux uncertainty is $\sim 0.08\%$.  The SAP flux is 
in units of count rate and the bandpass is from $4200$ \AA\ to $9000$ \AA. }
\label{fig:lc}
\end{figure}

The \textit{Kepler} light curves of the three AGNs are taken from \cite{Chen2015}. In their work, multiple-quarter 
light curves are stitched together by considering the PyKE routines \textit{kepmask} and \textit{kepextract} 
\citep{Kinemuchi2012}. Additional CBV (i.e., the cotrending basis vectors) corrections are not applied as such 
corrections are unlikely to be highly accurate at least for the best-studied source Zw 229-15 \citep[see Figure 
27 of][]{Smith2018}. The adopted light curves are presented in Figure~\ref{fig:lc}.

For each source, $M_{\mathrm{BH}}$ is fixed to the observed value; $\dot{M}$ is chosen to match the observed 
luminosity at rest-frame $\lambda=5100$ \AA; and $\delta_{\mathrm{mc}}$ is adjusted such that the predicted 
SF equals the observed one at $\Delta t=50$ days. The only remaining free parameter is $\alpha$ 
which determines the thermal timescale. We obtain mock light curves for two cases, i.e., $\alpha=0.01$ and 
$\alpha=0.2$. The former case (i.e., $\alpha=0.01$) corresponds to the results of some recent radiation MHD 
shearing box simulations \citep{Blaes2014}; the latter case (i.e., $\alpha=0.2$) is motivated by observational 
evidence \citep{King2007} from outbursts of dwarf nova or soft X-ray transients. 

We then solve Eq.~\ref{eq:energy} to obtain $T(t)$ using Euler's method. As a second step, we calculate the 
light curve of AGN UV/optical emission by assuming perfect blackbody radiation at each radius and a face-on 
viewing angle. To avoid sampling issues, the cadence of the mock light curve is 7-min which is higher than 
that of the \textit{Kepler} light curves. The duration of the mock light curve is $\sim 30$ years which is much 
(i.e., ten times) longer than that of the \textit{Kepler} light curves. 

To mimic the sampling patterns of the \textit{Kepler} light curves, we pick a segment of the mock light curve 
that has the same length and cadence as the observed light curve; the starting time of the segment is 
generated from a uniformly distributed random variable. We then add measurement noise to every segment 
by using uncorrelated white noise whose standard deviation is determined by the \textit{Kepler} observations. 
We repeat this process $512$ times (i.e., $512$ mock light curves, each with a duration of $\sim 30$ years, 
are generated) to account for the stochastic nature of the AGN UV/optical light curves. 

\begin{figure*}
\plotone{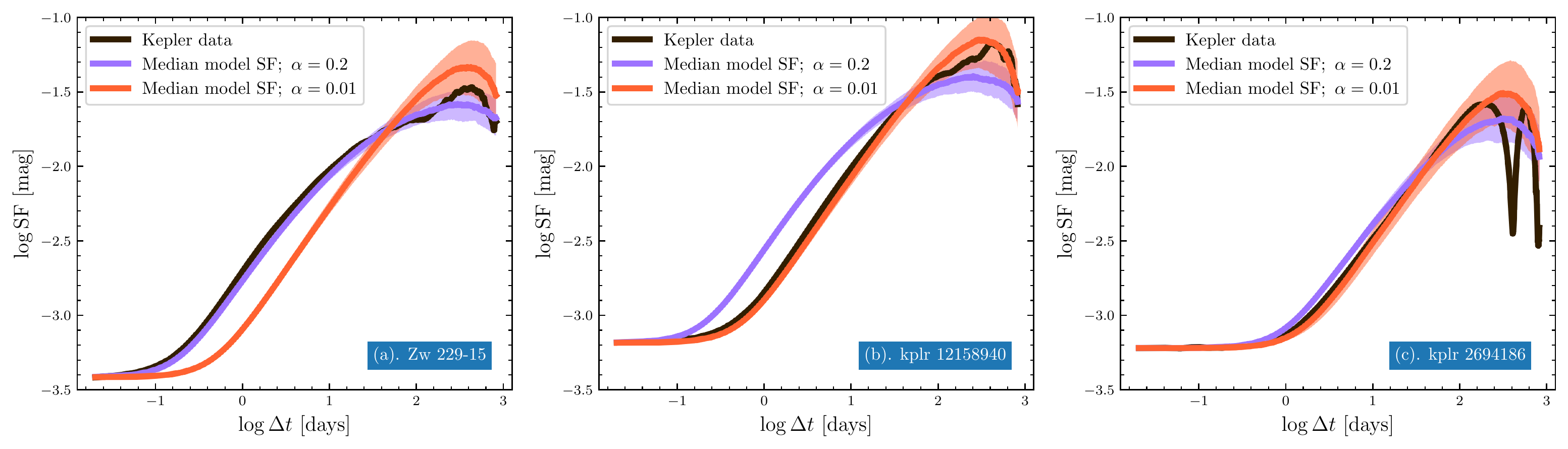}
\caption{The SFs for the three \textit{Kepler} AGNs. In each panel, the black thick curve 
represents the observed SF; the purple and orange thick curves correspond to our CHAR model 
with $\alpha=0.2$ and $\alpha=0.01$, respectively; the corresponding shaded regions indicate the 
$1\sigma$ uncertainties which account for the photometric noise and sampling effects. The structure 
functions show some dips or peaks at the long-timescale ends. This is simply because the light curve 
durations are too short to constrain the long-timescale variability. For kplr 2694186, the observed structure 
function has a dip feature around $\Delta t=400$ days which indicates periodicity in this source. However, 
this might be caused by instrumental effects \citep{Smith2018}. Note that when generating the model light 
curves, our simulations include the same time-sampling issues and photometric errors as the structure 
functions calculated from the real data. }
\label{fig:sf}
\end{figure*}

We compute the SFs of the observed and simulated light curves. For each AGN, a set of $512$ 
simulated SFs can be obtained. Following \cite{Kasliwal2015}, we calculate the ensemble mean 
$\overline{\log_{10} \mathrm{SF}(\Delta t)}$ and standard deviation $\sigma_{\mathrm{SF}}$ of the set of $512$ 
SFs. The SFs of our mock light curves for the three sources are presented in 
Figure~\ref{fig:sf}. The similarities between our model SFs and the observed ones on short timescales 
(which cover nearly two orders of magnitude in timescale, i.e., from $\gtrsim 0.5$ days up to $\lesssim 50$ days) 
are intriguing (see Figure~\ref{fig:sf}) since the shapes of the model SFs on short timescales are 
primarily determined by the thermal-energy conservation law of the accretion disk ($\alpha$ has limited impact on 
the short-timescale SFs). This result indicates that our CHAR model reveals the physical nature of the disk 
temperature fluctuations. In other words, with appropriate $\alpha$ values, our CHAR model can almost 
precisely reproduce 
the observed SFs on all covered timescales for all the three Kepler AGNs with extremely high-cadence 
and long-duration light curves (i.e., the best optical AGN light curves ever in terms of these two aspects). 

We use the following pseudo $\chi^2$ statistic to assess quality-of-fit,
\begin{equation}
\label{eq:chi2}
\chi^2_{\mathrm{obs}} = \sum \frac{(\overline{\log_{10} \mathrm{SF}(\Delta t)}-\log_{10} 
\mathrm{SF}_{\mathrm{obs}}(\Delta t))^2}{\sigma_{\mathrm{SF}}^2} \\,
\end{equation}
where $\mathrm{SF}_{\mathrm{obs}}(\Delta t)$ is the SF of an observed light curve. A list of the 
ratio of $\chi^2$ to the degrees of freedom (D.O.F.) can be found in Table~\ref{tbl:tb2}. 

The pseudo $\chi^2$ statistic does not follow the classical $\chi^2$ distribution because the adjacent 
SF estimates are correlated and for other statistical reasons \citep{Emmanoulopoulos2010}. To 
assess the quality-of-fit, we must use simulations to obtain the distribution of our pseudo $\chi^2$ 
\citep{Uttley2002,Kasliwal2015}. That is, we use Eq.~\ref{eq:chi2} to obtain the pseudo $\chi^2$ (hereafter 
$\chi^2_{\mathrm{mc}}$) for each of the $512$ simulated SFs; in this step, 
$\mathrm{SF}_{\mathrm{obs}}$ is replaced with the simulated SF. The distribution of $512$ 
$\chi^2_{\mathrm{mc}}$ can be used to infer the distribution of the pseudo $\chi^2$ for our CHAR model. We then 
define a new statistical quantity, the likelihood of occurrence ($P(\chi^2_{\mathrm{mc}}>x)$), which measures 
the probability of $\chi^2_{\mathrm{mc}}$ taking a value larger than a specific value $x$ (i.e., statistically 
speaking, $P(\chi^2_{\mathrm{mc}}>x)$ is the survival function of the distribution of $\chi^2_{\mathrm{mc}}$). 

\begin{figure*}
\plotone{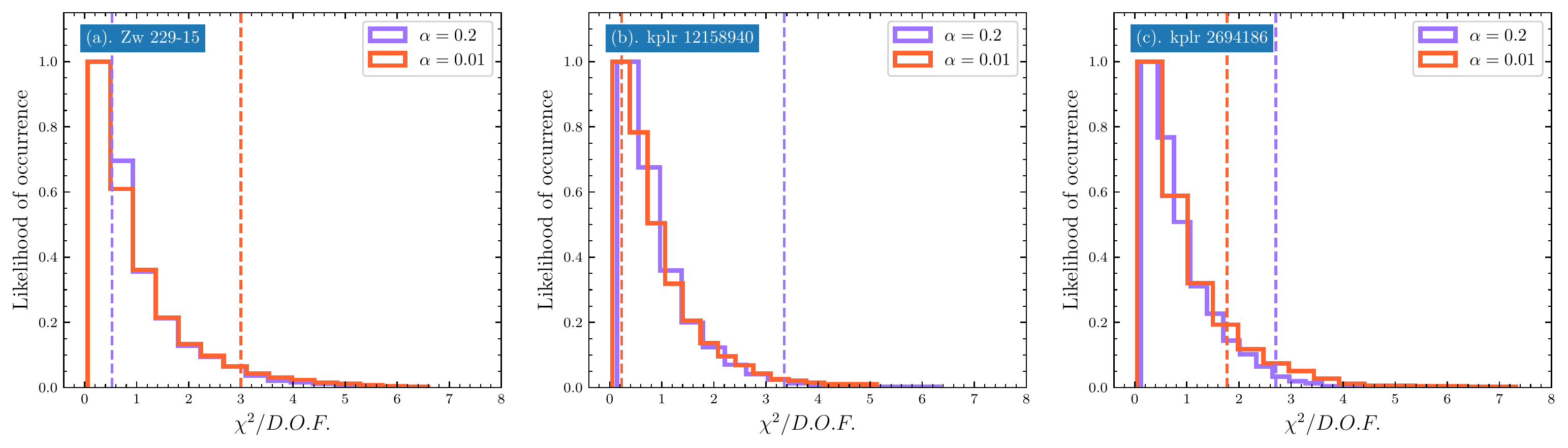}
\caption{The likelihood of occurrence ($P(\chi^2_{\mathrm{mc}}>x)$), which measures the probability 
of $\chi^2_{\mathrm{mc}}$ taking a value larger than a specific value $x$ (i.e., statistically speaking, 
$P(\chi^2_{\mathrm{mc}}>x)$ is the survival function of the distribution of $\chi^2_{\mathrm{mc}}$). 
In each panel, the purple and orange vertical lines indicate the ratio of $\chi^2_{\mathrm{obs}}$ to 
D.O.F. for our CHAR model with $\alpha=0.2$ and $\alpha=0.01$, respectively. $P(\chi^2_{\mathrm{mc}}> 
\chi^2_{\mathrm{obs}})$, i.e., the likelihood of occurrence at the crosspoint between the dashed line 
and the solid histogram of each color, approaching $0$ indicates that the model is a poor fit to the data. }
\label{fig:gdness}
\end{figure*}

The likelihood of occurrence of each source is shown in Figure~\ref{fig:gdness}. For comparison, we also show 
$\chi^2_{\mathrm{obs}}$ for each source. For Zw 229-15, our CHAR model with $\alpha=0.01$ is a poor fit (the fit is 
even poorer if we focus only on $\Delta t<50$ days) since $P(\chi^2_{\mathrm{mc}}>\chi^2_{\mathrm{obs}})$ 
is $0.05$; and the model with $\alpha=0.2$ is a good fit because $P(\chi^2_{\mathrm{mc}}>\chi^2_{\mathrm{obs}}) 
\sim 0.67$. For kplr 12158940, our CHAR model with $\alpha=0.2$ is a poor fit since $P(\chi^2_{\mathrm{mc}}> 
\chi^2_{\mathrm{obs}})$ is $0.02$; instead, the model with $\alpha=0.01$ is a reasonable fit because 
$P(\chi^2_{\mathrm{mc}}>\chi^2_{\mathrm{obs}}) = 0.92$. For the same reason, our CHAR model with $\alpha=0.01$ 
($P(\chi^2_{\mathrm{mc}}>\chi^2_{\mathrm{obs}})=0.15$) is a better fit to the observed SF of 
kplr 2694186 than that with $\alpha=0.2$ ($P(\chi^2_{\mathrm{mc}}>\chi^2_{\mathrm{obs}})=0.03$). 

\begin{figure}
\plotone{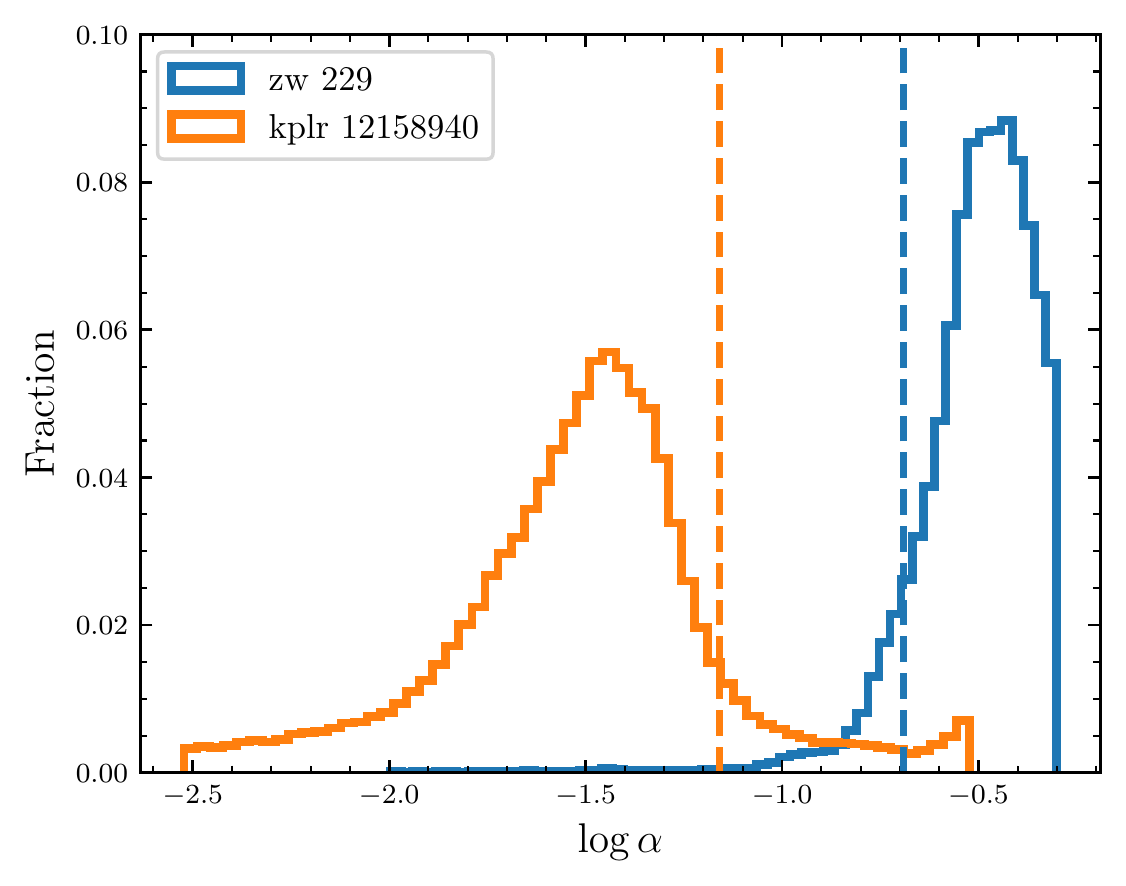}
\caption{The probability distributions of $\alpha$ for Zw 229-15 (blue curve) and kplr12158940 (orange 
curve). The vertical dashed blue and orange lines indicate the $10$-th and $90$-th percentiles of the distributions 
of $\alpha$ for Zw 229-15 and kplr12158940, respectively. That is, the probability that the required $\alpha$ for 
Zw 229-15 is smaller than that for kplr12158940 is less than $10\% \times 10\%=1\%$.} 
\label{fig:alpha_mcmc}
\end{figure}

To demonstrate the statistical distribution of $\alpha$, we perform the following calculations. 
For Zw 229-15 and kplr12158940 (we exclude kplr 2694186 because the model with $\alpha=0.01$ is only 
slightly better than that with $\alpha=0.2$), our model SFs are calculated by stepping through sixteen values of 
$\alpha$ from $0.01$ to $0.5$ in equal logarithmic increments. The likelihood that the observed SF is a 
realization of our model with a specified $\alpha$ is estimated by considering the pseudo $\chi^2$ and the 
distribution of 512 $\chi^2_{\mathrm{mc}}$ as outlined above. For each source, we then interpolate 
the sixteen likelihoods to estimate the likelihoods of other values of $\alpha$ and adopt the popular \textit{Python} 
implementation of the Markov Chain Monte Carlo algorithm, \textit{emcee} \citep{emcee}, to sample the model 
parameter $\alpha$. The resulting distributions of $\alpha$ for Zw 229-15 and kplr12158940 are shown in 
Figure~\ref{fig:alpha_mcmc}. The required $\alpha$ for Zw 229-15 is indeed statistically larger than that for 
kplr12158940 since the possibility that the required $\alpha$ for Zw 229-15 is smaller than that for kplr12158940 
is less than $1\%$. 

Our results demonstrate that, for light curves with sufficient quality (especially on long timescales), we can in 
principle infer the value of $\alpha$ by fitting the AGN UV/optical light curves. The fits to the three Kepler AGNs 
already suggest that different AGNs have different $\alpha$ values. The required $\alpha$ is not entirely 
consistent with the values derived from some recent radiation MHD shearing box simulations of accretion disks 
where $\alpha$ converges around $0.01$ \citep{Blaes2014}. A similar discrepancy is also found when analyzing 
the observations of outbursts of dwarf nova or soft X-ray transients \citep{King2007}. Some possible 
explanation involve the large-scale poloidal magnetic field (as illustrated in Figure~\ref{fig:illu}, the large-scale 
magnetic field is also required in our CHAR model) because $\alpha$ positively correlates with the initial field 
strength \citep[see Figure 6 of][]{Hawley1995} or the kinetic effect of MRI turbulence \citep[e.g.,][]{Kunz2016}; 
a detail discussion of additional possibilities has been made by \cite{King2007}.

\subsection{NGC 5548}
\label{sect:ngc5548}
Our CHAR model also has the potential to self-consistently account for other observational characteristics of AGN 
UV/optical variability, e.g., the multi-wavelength variability of the best-studied reverberation-mapped AGN, 
NGC 5548. To apply our CHAR model to NGC 5548, we fix the mass of the central SMBH to be $M_{\mathrm{BH}} 
=5\times 10^7\ M_{\odot}$ and choose $\dot{M}$ such that the model luminosity at $5100$ \AA\ is consistent 
with the observed one \citep{Fausnaugh2016}; $\delta_{\mathrm{mc}}$ is adjusted to ensure that the structure 
function at $10$ days of the model light curve at $B$-band matches the observed one. The remaining 
parameter is $\alpha=0.2$. We then run simulations to generate eighteen-band model light curves (i.e., all 
the eighteen UV/optical bands listed in table~5 of \citealt{Fausnaugh2016}; we do not consider X-ray 
observations because X-ray emission is not produced by the accretion disk but by the hot corona) following 
the methodology mentioned above. During the simulations, the time-sampling issues and the measurement 
errors are also considered, i.e., the model light curves share the same cadence and measurement 
noise as the observations of NGC 5548 \citep{Fausnaugh2016}. 

\begin{figure}
\plotone{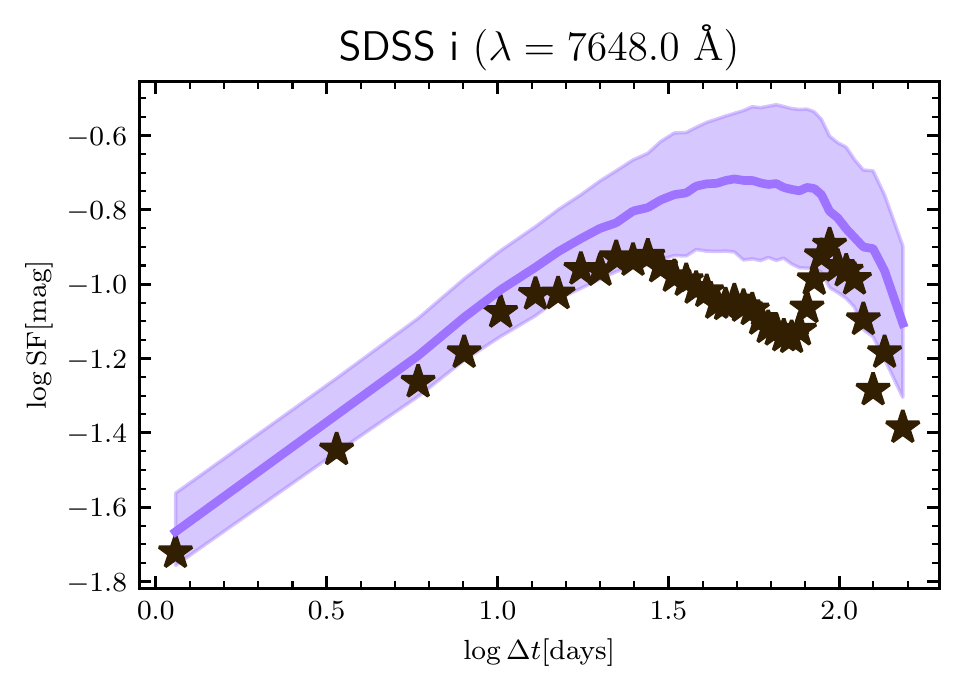}
\caption{The SFs of NGC 5548 for various wavelengths (annotated top). The stars and the purple 
curve represent the observed and the model SFs, respectively; the shaded regions indicate the 
$1\sigma$ uncertainties. The SFs for all eighteen bands are available online. }
\label{fig:sf-5548}
\end{figure}

\begin{figure}
\plotone{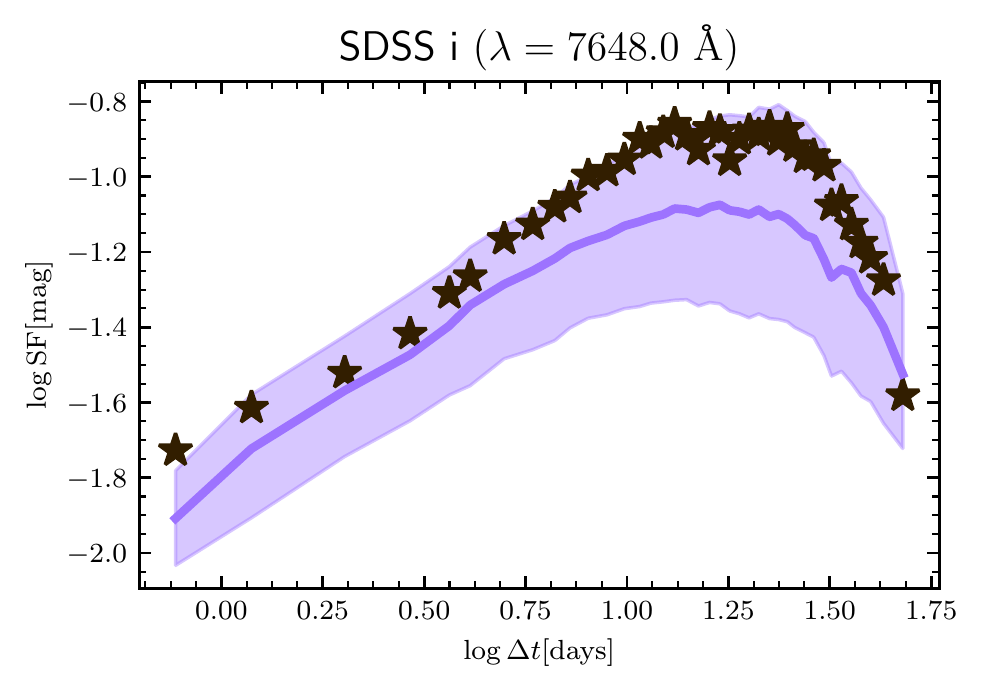}
\caption{The SFs of NGC 5548 for various wavelengths (annotated top). Unlike 
Figure~\ref{fig:sf-5548}, only the first portions (i.e., data points with $\mathrm{HJD}-2450000<6747$) 
of the light curves are considered here. The stars and the purple curve represent the observed and 
the model SFs, respectively; the shaded regions indicate the $1\sigma$ uncertainties. 
The SFs for all eighteen bands are available online. }
\label{fig:sf-5548-gd}
\end{figure}

For each band, we first compare the model SF with the observed one (see Figure~\ref{fig:sf-5548} 
for the SDSS $i$-band; the complete figure set for all bands is available online). Overall, our 
model can account for the observed SFs of the multi-band light curves of NGC 5548 on timescales 
$\lesssim 20$ days. On timescales of $20$--$50$ days, our CHAR model over-predicts the observed 
variability. This deviation might have something to do with the anomalous state \citep{Goad2016} in 
NGC 5548; in the anomalous state, the ionizing continuum is preferentially suppressed 
due to, e.g., the intrinsic change of the corona/disk structure \citep{Mathur2017,Sun2018a} or external 
variations in line-of-sight obscuration \citep{Dehghanian2019, Kriss2019}. Indeed, if we split the full 
multi-wavelength light curves of NGC 5548 into two segments at $\mathrm{HJD}-2450000<6747$, the 
first portions are more variable than the second ones, especially on timescales longer than $10$ days 
\citep[see Figure~5 of][]{Sun2018a}. We then re-apply our CHAR model to the first segment of each 
band of the NGC 5548 light curve following the same methodology. The resulting SFs are shown in 
Figure~\ref{fig:sf-5548-gd}. Our CHAR model can now also explain the observed SFs of NGC 5548 
on timescales $\gtrsim 20$ days. Therefore, our results indicate that the magnetic fluctuations in the 
corona might also change as NGC 5548 entered into the anomalous state. 

\begin{figure}
\plotone{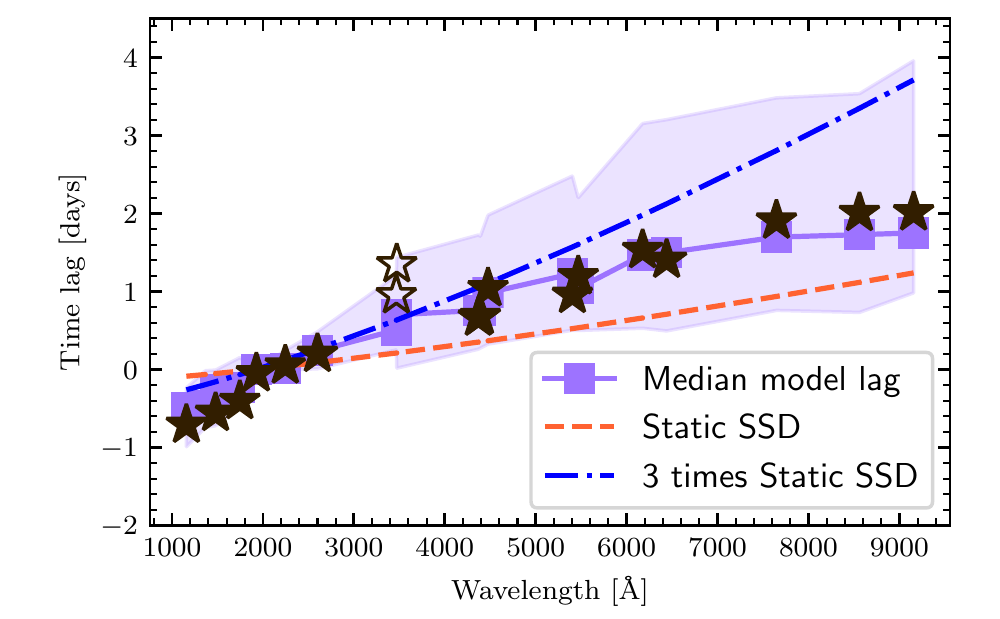}
\caption{The inter-band time lags of NGC 5548. Only the first portions (i.e., data points with $\mathrm{HJD}-2450000 
<6747$) of the light curves are considered here. The stars and the squares represent the observed and the median 
model time lags, respectively. The open stars correspond to the time lags of the $U$ and $u$ bands which are likely 
to be affected by broad-line region emission. The purple shaded regions indicate the $1\sigma$ uncertainties of 
the model time lags. The red dashed curve shows the expected time lags of the static thin disk model; the blue 
dashed-dot curve indicates the expected time lags if we increase the static thin disk sizes by a factor of three. Note 
that our modeling approach for NGC 5548 includes the same time-sampling pattern and photometric errors as the 
inter-band time lags estimated from the real light curves. }
\label{fig:tlag-5548}
\end{figure}

We then use \textit{PYCCF} \citep{PYCCF}, a python version of the interpolation cross-correlation function code 
\citep{Peterson1998}, to determine the inter-band time lags for the model light curves; the reference band is 
chosen to be the \textit{Swift} UVW2 band \citep{Edelson2019}. We also use our code to re-estimate the observed 
time lags for the sake of self-consistency. We limit our analyses to the first segments of the multi-wavelength light 
curves of NGC 5548. Our CHAR model can explain the observed inter-band time lags which are larger than the 
expectations of the lamppost model (see Figure~\ref{fig:tlag-5548}). This is because, unlike the lamppost model, 
the disk temperature cannot fully respond to the variations of $Q^{+}_{\mathrm{heat}}$ unless a thermal timescale 
has passed. That is, the inter-band time lags are superpositions of the magnetic fluctuation travel (the speed is 
assumed to be the speed of light) timescales and the response timescales (see Section~\ref{sect:ccf}). 

\section{Discussion}
\label{sect:dis}
\subsection{Physical Mechanisms}
\label{sect:dis1}
Considering the large optical depth from the surface to the mid-plane, external illumination 
(e.g., X-ray or UV emission) should be absorbed within a thin surface of the accretion disk. If 
so, the timescale for such a thin surface to adjust its structure and the absorbed energy to be 
reprocessed as UV/optical emission is rather short \citep[it can be less than one hour; see, 
e.g.,][]{Collin1991, Czerny2006}; that is, Eq.~\ref{eq:qqq} seems to be valid. However, as we 
discussed in Section~\ref{sect:intro}, this simple and interesting model fails to explain many 
observational facts of AGN UV/optical variability \citep[see also][]{Edelson2019}. 

To overcome these problems, we propose that the corona and the accretion disk are tightly 
coupled by magnetic fields. Turbulent magnetic fields are well expected in an accretion disk  
since the MRI is responsible for removing angular momentum, releasing the gravitational energy 
and heating the gas in the accretion disk \citep{Balbus1998}. The interior magnetic 
fields might rise to the low-density surface owing to, e.g., magnetic buoyancy \citep{Parker1966} 
and can be effectively amplified to form large-scale poloidal magnetic fields \citep{Rothstein2008}. 
In the vertical regions that are well above the accretion disk, the gas is highly magnetized; the 
puffed-up magnetic field might reconnect, dissipate its energy, and heat the ambient low-density 
plasma \citep[e.g.,][]{DiMatteo1998, Liu2002}. If the dissipated energy is mainly converted into the 
internal energy of protons, and protons and electrons are largely decoupled (given the low-density 
nature, Coulomb coupling is inefficient; see e.g., \citealt{DiMatteo1998, Rozanska2000}), the plasma 
will be radiatively inefficient \citep[which is similar to an advection-dominated accretion flow; see][]{adaf}. 
This hot and radiatively inefficient plasma might be responsible for the so-called ``X-ray corona''. 
The magnetic power might also launch a relativistic jet from the coronal plasma \citep[i.e., the corona 
might serve as the jet base;][]{Markoff2005}. In addition, the SMBHs might be supplied by the ambient 
hot gas (e.g., from stellar winds) and the accretion flow around the Bondi radius can be geometrically 
thick; an underlying thin and cold disk only forms at much smaller radii; then, the ``X-ray corona'' might 
be the innermost regions of this thick disk \citep{Liu2015}. Unlike the underlying cold thin disk \citep{SSD}, 
the corona should have a large inflow velocity \citep{Liu2015,Jiang2019} since the plasma is hot and 
the angular-momentum transfer due to the MRI should be efficient. Therefore, the anchored magnetic 
field in the corona can be ``dragged'' into the innermost regions. The same 
magnetic field that penetrates the interior of the cold accretion disk remains in its original radial 
location as the inflow velocity of the cold disk is small. Therefore, a magnetic coupling between 
the compact corona and the outer cold accretion disk might exist (see Figure~\ref{fig:illu}).

As the magnetic field of the corona fluctuates (due to, e.g., magnetic reconnection), the disk turbulent 
magnetic field also changes accordingly after a time delay which accounts for the propagation of MHD 
waves from the corona to the disk; the time delay $t_{\mathrm{delay}}$ is $R_X/c_{\mathrm{avf}}$, 
where $R_X$ and $c_{\mathrm{avf}}$ are the distance between the corona and the disk and the Alfv\'en 
velocity, respectively. The coherently variable disk turbulent magnetic power (i.e., the fluctuations of disk 
turbulent magnetic field at different radii are correlated) dissipates and changes the heating rate in the 
disk. As a result, the interior structure of the accretion disk changes in response to the variable disk 
heating rate. The timescale for the disk temperature to adjust to the variable disk heating rate is the 
thermal timescale. On timescales significantly longer than the thermal timescale, the 
disk temperature and the disk heating rate vary similarly; but on timescales shorter than the 
thermal timescale, the disk response time is important and the disk variability is less than the 
fluctuation in the heating rate. This naturally leads to less variability on short timescales and more 
variability on long timescales, explaining why the thermal timescale is a good fit to the ``break'' 
timescale between the two variability regimes \citep[e.g.,][]{Kelly2009,Sun2015,Sun2018c}.

\subsection{The Correlation Between X-ray and UV/optical Variability}
\label{sect:dis2}
In this work, we assume that the disk-temperature variations are induced by corona magnetic 
fluctuations; the same magnetic fluctuations can also drive X-ray variability. Therefore, one might 
expect a tight correlation between UV/optical and X-ray emission. However, the relationship between 
\qill\ and X-ray emission can be complicated due to the important advective cooling and the 
fluctuations of the corona surface density (see also Section~\ref{sect:ccf}). A detailed investigation of 
this topic is needed to understand the relation between X-ray and UV/optical stochastic variations; 
however, this is beyond the scope of this work.

\subsection{Relationship To Other Models}
\label{sect:dis3}
Alternative models have been proposed to explain AGN UV/optical light curves. One of the most 
popular models is the X-ray reprocessing model \citep[e.g.,][]{Krolik1991}. In this model, 
the highly variable X-ray emission (which is presumably produced in the hot corona) can illuminate 
the underlying cold accretion disk; a significant fraction of the illuminating X-ray photons are 
thermalized in the disk surface. The absorbed X-ray emission is reprocessed as the UV/optical 
emission which might be responsible for the observed AGN UV/optical light curves. However, this 
scenario is challenged by many observations (\citealt{Uttley2003, SunYH2014,Fausnaugh2016, 
Cai2018,Kang2018,Zhu2018,Edelson2019}; see also Section~\ref{sect:intro}). 

Variations of accretion rate at each radius can also induce AGN luminosity fluctuations. In fact, 
\cite{Lyubarskii1997} demonstrated that, if the accretion rate at each radius varies independently, 
the PSD of the AGN bolometric luminosity is $\propto 1/f$. However, the required timescale for 
the accretion rate to vary is the viscous timescale, which should be around hundreds of years for the 
UV/optical emission regions of a typical AGN. Therefore, this model cannot explain the observed 
UV/optical variability. Instead, it might be able to explain the short-timescale (i.e., hours to years) 
magnetic energy fluctuations in the innermost regions or the compact corona where the 
corresponding viscous timescale can be less than days. 

Another popular model is the strongly inhomogeneous disk model \citep{Dexter2011,Cai2016}. 
While this model has the potential to explain the microlensing observations \citep{Morgan2010}, 
the timescale-dependent AGN color variability \citep{SunYH2014}, and many other observational 
characteristics, the temperature fluctuations in this model are ``assumed'' to be a DRW process. 
Meanwhile, this model fails to explain the inter-band cross correlations since the temperature 
fluctuations at different radii vary independently. \cite{Cai2018} upgraded the strongly 
inhomogeneous disk model by adding a global common temperature fluctuation and found that 
this new model has the potential to yield the observed inter-band UV/optical time 
lags \citep{Fausnaugh2016,Edelson2019}. However, temperature fluctuations in \cite{Cai2018} 
are still assumed to be a DRW process. 

All in all, a model to account for the AGN UV/optical variability has long been lacking until our 
work where we propose a simple way to determine the temperature fluctuations in the accretion 
disk and explain AGN UV/optical light curves.

\subsection{LSST Forecasts}
\label{sect:lsst}
It has been shown in previous sections that our CHAR model can well explain the multi-wavelength variability 
of the AGNs that have the best-quality observations available . The variability properties depend critically 
on our CHAR model's free parameters, namely, $M_{\mathrm{BH}}$, $\alpha$, $\dot{M}$ (or $\dot{m}$) and 
$\delta_{\mathrm{mc}}$. It is then possible to constrain these parameters by adopting our CHAR model to 
fit future LSST light curves. To illustrate this idea, we consider five AGNs with five 
different choices of $M_{\mathrm{BH}}$, $\dot{M}$, and $\alpha$, i.e., $M_{7}=1$, $\dot{M}_{24}=1.3$, 
and $\alpha=0.2$ (hereafter case I); $M_{7}=5$, $\dot{M}_{24}=1.3$, and $\alpha=0.2$ (hereafter 
case II); $M_{7}=5$, $\dot{M}_{24}=0.08$, and $\alpha=0.05$ (hereafter case III); $M_{7}=1$, 
$\dot{M}_{24}=6.5$, and $\alpha=0.2$ (hereafter case IV); $M_{7}=5$, $\dot{M}_{24}=1.3$, and 
$\alpha=0.01$ (hereafter case V). Note that $M_7=M_{\mathrm{BH}}/(10^7 M_{\odot})$ and 
$\dot{M}_{24}=\dot{M}/(10^{24}\ \mathrm{g\ s^{-1}})$. 

For each case, we use our CHAR model to simulate the light curves of the observed-frame $3500$ \AA\ and 
$8500$ \AA\ (which correspond to the central wavelengths of the $u$ and $z$ bands of the LSST filters, 
respectively) emission; the duration of every light curve is $10$ years (in the observed frame); the 
photometric noise is assumed to be $0.01$ mag and the cadence of the simulations is (observed-frame) 
$3$-day which is motivated by the LSST surveys of the deep-drilling fields \citep{Brandt2018,Scolnic2018}. 
For each case, we repeat the simulation $512$ times to account for statistical fluctuations due to 
photometric noise, limited cadence, and duration. For each case, $\delta_{\mathrm{mc}}$ is chosen to 
ensure that the SF of the $3500$ \AA\ emission at $50$ days is the same (i.e., $\cong 
0.03$ mag). 

The SFs of the observed-frame $3500$ \AA\ and $8500$ \AA\ emission and their ratios are calculated. The 
ratios are similar to the color variability in Section~\ref{sect:color}. That is, a bluer-when-brighter 
behavior is expected if the ratio is smaller than unity. 

\begin{figure*}
\epsscale{1.2}
\plotone{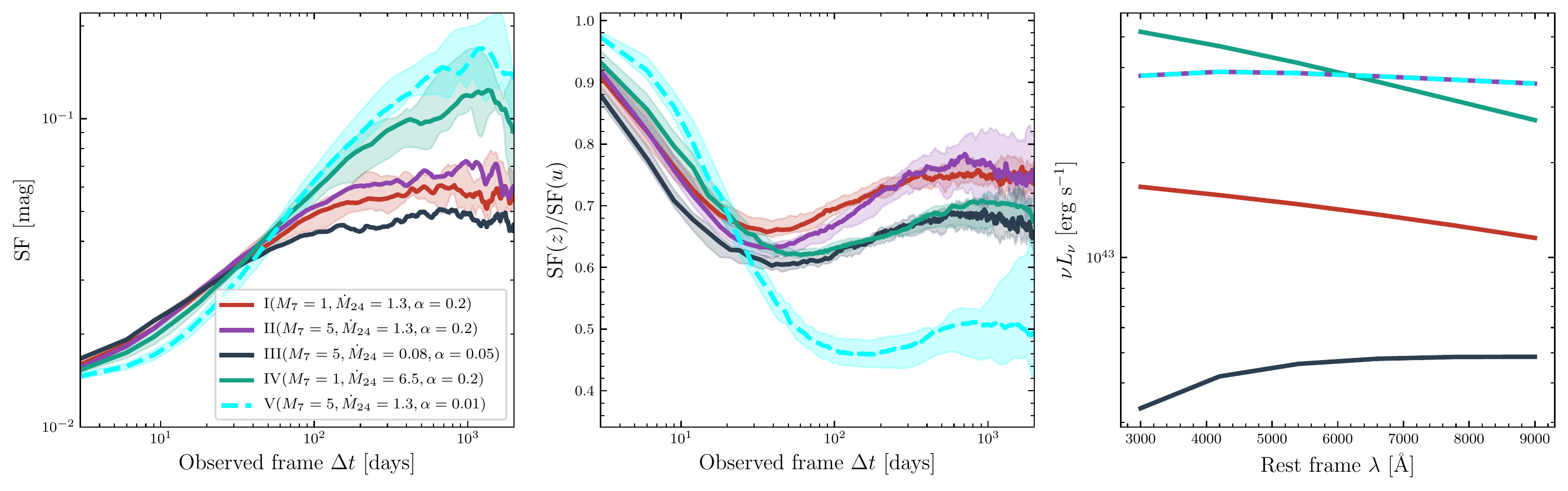}
\caption{The predicted statistical properties of AGN UV/optical light curves as a function of $M_7$, $\dot{M}_{24}$ 
and $\alpha$, where $M_7=M_{\mathrm{BH}}/(10^7 M_{\odot})$ and $\dot{M}_{24}=\dot{M}/(10^{24}\ \mathrm{g\ s^{-1}})$. 
The cadence is assumed to be 3 days which is motivated by the LSST surveys in the deep-drilling fields. Left: 
the SFs of the observed-frame $3500$ \AA\ emission (which corresponds to the LSST $u$ band). 
Middle: the ratios of the SFs of the $8500$ \AA\ emission (which corresponds to the LSST 
$z$ band) to those of the observed-frame $3500$ \AA\ emission (the ratios are similar to the color variability 
in Section~\ref{sect:color}, i.e., a bluer-when-brighter behavior is expected if the ratio is smaller than unity). 
Right: the corresponding spectral energy distributions. The shaded regions represent the $1\sigma$ 
uncertainties (the uncertainties of the SFs of cases II and III are not shown for the purpose of clarity). 
By simultaneously considering all the three properties, we can distinguish AGNs with different physical parameters 
and thereby constrain $M_{\mathrm{BH}}$, $\dot{M}$, and $\alpha$ given LSST datasets. The redshifts of these 
mock AGNs are fixed to $z=0.017175$ (i.e., the same as that of NGC 5548). }
\label{fig:lsst}
\end{figure*}

\begin{figure*}
\epsscale{1.2}
\plotone{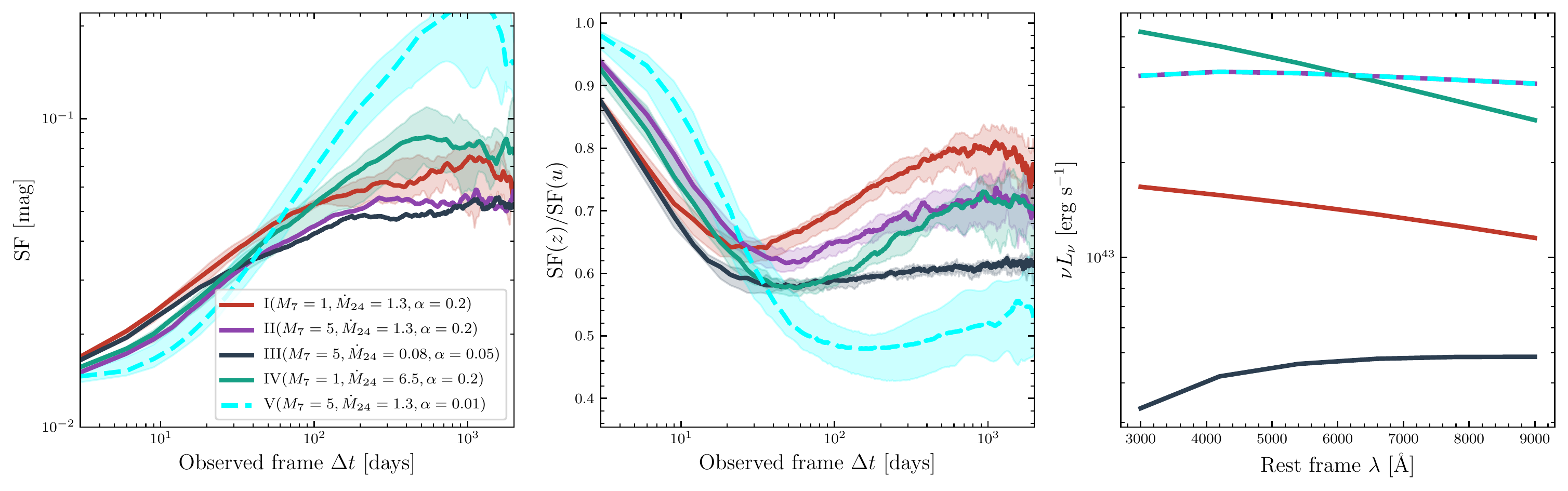}
\caption{Same as Figure~\ref{fig:lsst}, but for mock AGNs at $z=1$. }
\label{fig:lsst-z1}
\end{figure*}

The SFs of the $3500$ \AA\ emission, the ratios of the SFs of the $8500$ \AA\ emission to 
those of the $3500$ \AA\ emission, and the expected SEDs are shown in Figure~\ref{fig:lsst} 
for AGNs at $z=0.017175$ (i.e., the same as that of NGC 5548). The results for the same 
AGNs at $z=1$ are presented in Figure~\ref{fig:lsst-z1}. The differences in the SFs are evident 
beyond the measurement noise if the corresponding $\tau_{\mathrm{TH}}$ ($\propto 
\dot{M}^{0.5}\alpha^{-1}$; see Eq.~\ref{eq:scale}) values are significantly different (i.e., comparing 
case I or II with case IV or V). While the SFs of cases A, B and C (they have similar 
$\tau_{\mathrm{TH}}$) are indistinguishable within measurement noise, their color 
variability is statistically different. Therefore, it is promising to infer $\dot{M}$ and $\alpha$ by 
considering the LSST light curves of thousands of type I AGNs in the LSST deep-drilling 
fields. The SEDs are sensitive to both $\dot{M}$ and $M_{\mathrm{BH}}$. Hence, as long as 
$\dot{M}$ is determined, we can also infer $M_{\mathrm{BH}}$ from the LSST data. Note that 
many of the brightest AGNs in the LSST deep-drilling fields will also have \textit{independent} 
$M_{\mathrm{BH}}$ measurements from the reverberation-mapping campaigns, e.g., the 
SDSS-V Black Hole Mapper \citep{Kollmeier2017} and 4MOST/TiDES \citep{Swann2019}. 
Then, we can use the two independent $M_{\mathrm{BH}}$ measurements to perform a 
cross-validation study to improve the accuracy of $M_{\mathrm{BH}}$. Moreover, the radiative efficiency 
$\eta \equiv L_{\mathrm{bol}}/(\dot{M}c^2)$ can be calculated. For radiatively efficient accretion 
disks, $\eta$ should be determined by the innermost stable circular orbit (ISCO) radius 
$R_{\mathrm{ISCO}}$ and magnetic stress at this radius 
\citep[][]{Agol2000}; $R_{\mathrm{ISCO}}$ depends on the SMBH spin $a^{*}$. Therefore, we can 
use the measured $\eta$ to deliver some insights on $a^{*}$ and/or magnetic stress at 
$R_{\mathrm{ISCO}}$.

\section{Summary}
\label{sect:sum}
We propose a new model, Corona-Heated Accretion-disk Reprocessing (a.k.a., CHAR), 
to explain AGN UV/optical variability. In contrast to the simplest X-ray reprocessing 
model, we argue that, as the corona induces fluctuations in the heating rate, the temperature 
of the interior of an AGN accretion disk also changes. We assume that the AGN accretion 
disk can re-establish vertical hydrostatic equilibrium, and the variability of \qill\ can be 
described by a red-noise process. Then, the temperature fluctuations can be determined 
by considering the vertically integrated thermal-energy conservation law (see 
Eq.~\ref{eq:Ttime} and Section~\ref{sect:model}). 

We solve Eq.~\ref{eq:Ttime} to obtain the temperature fluctuations and luminosity 
variability. We find that the fluctuations of the inner and surface temperature and luminosity 
differ from that of \qill\ in many aspects. Our main results can be 
summarized as follows. 
\begin{enumerate}
\item The fluctuations of the inner and surface temperature and luminosity contain less 
high-frequency components than that of \qill\ (see Figures~\ref{fig:lc_A}, 
\ref{fig:TPSD} and \ref{fig:sf_long}; Section~\ref{sect:ssp}). 

\item According to our CHAR model, on timescales of $10^2$---$10^3$ days, AGN UV/optical 
luminosity variability can be well fitted by the DRW process; on shorter/longer timescales, the 
DRW process under-predicts/over-predicts AGN UV/optical luminosity variability (see 
Figure~\ref{fig:sf_long}). 

\item The PSD and SF of AGN UV/optical luminosity variability have a characteristic timescale, 
i.e., the thermal timescale $\tau_{\mathrm{TH}}$ (see Eq.~\ref{eq:scale}; Section~\ref{sect:result_mk}).

\item The PSD and SF depend mostly on AGN luminosity or $\dot{M}$ (see Figure~\ref{fig:PMS_ALL}; 
Section~\ref{sect:pms}); their dependences on $M_{\mathrm{BH}}$ or $\dot{m}$ are 
weak. 

\item AGN UV/optical luminosity variability decreases with increasing wavelength; the difference is more 
evident on short timescales (see Figure~\ref{fig:wave_A}). 

\item Our CHAR model predicts a bluer-when-brighter behavior (see Figure~\ref{fig:diff}); the 
bluer-when-brighter behavior is more evident on short timescales than on long timescales 
(see Figure~\ref{fig:vcolor}). 

\item AGN timescale-dependent color variability (bluer-when-brighter) correlates with $\dot{m}$ and 
$L_{\mathrm{bol}}$; its dependence on $M_{\mathrm{BH}}$ is complex (see Figures~\ref{fig:diff} and 
\ref{fig:vcolor}; Section~\ref{sect:mbh}). 

\item Unlike the X-ray reprocessing model, the inter-band time lags of our CHAR model increase with 
increasing timescales (see Figure~\ref{fig:tlag_A}; 
Section~\ref{sect:ccf}). For an AGN with $M_{\mathrm{BH}}=10^8\ 
M_{\odot}$ and $L=0.1 L_{\mathrm{Edd}}$, on timescales of $\sim 10^2$ days, the inter-band 
time lags between UV and optical bands can be $\sim 3$ times larger than the expectations of 
the static SSD model. 

\item Our CHAR model might also be able to explain AGN microlensing observations (see Figure~\ref{fig:rhalf}; 
Section~\ref{sect:micro}). 

\item Our CHAR model can successfully explain the high-quality Kepler AGN light curves 
(see Figure~\ref{fig:sf}; Section~\ref{sect:kepler}); the dimensionless viscosity, one of 
the basic parameters in the black hole accretion theory, which cannot be determined 
by fitting AGN SEDs, is constrained to be $0.01$---$0.2$ by our CHAR model. 

\item Our CHAR model can also account for the lager-than-expected time lags in NGC 5548 
(see Figure~\ref{fig:tlag-5548}). With the same parameters, our CHAR model can simultaneously 
fit the SFs of the eighteen light curves of NGC 5548 (see Figure~\ref{fig:sf-5548-gd}; 
Section~\ref{sect:ngc5548}). 

\item We demonstrate that $M_{\mathrm{BH}}$, $\dot{M}$, and $\alpha$ can be constrained 
by applying our CHAR model to fit AGN multi-band light curves from LSST time-domain surveys 
(see Figures~\ref{fig:lsst} and \ref{fig:lsst-z1}; Section~\ref{sect:lsst}).
\end{enumerate}

Therefore, our CHAR model has the potential to explain many observational facts about AGN UV/optical 
variability. 

If our CHAR model is correct, the time lag between optical and the ionizing continuum emission can be 
significant on long timescales (see Figure~\ref{fig:tlag_A}; Section~\ref{sect:ccf}). Most RM campaigns 
usually measure the time lag between the BEL and the nearby optical emission. Therefore, the distance 
of BLR to the central SMBH can be significantly under-estimated 
for a long-term (i.e., the nearby continuum light curve contains long-term variability) RM campaign. 
This bias can be corrected by performing long-term detrending to the RM light curves. 

Our work can be advanced in some theoretical aspects. For instance, disk winds can be strong and 
modify the structure of the accretion disk \citep{Sun2019} and the disk emission may not be a perfect 
black body \citep{Hall2018}. We also ignore the UV/optical variability due to X-ray reprocessing of 
a static SSD or diffuse BLR clouds \citep[e.g.,][]{Cackett2018, Sun2018a}. It would be interesting to 
revise our CHAR model to include these physical processes. In addition, our analysis cannot be applied 
to timescales comparable to the viscous 
timescales unless accretion-rate fluctuations (which can be significant due to, e.g., radiation-pressure 
instabilities) are properly modeled; such accretion-rate fluctuations have been proposed to explain 
the intermittent activity of compact GPS radio sources \citep{Czerny2009}.

In future works, we will test our model with additional observations, e.g., the inter-band cross correlations 
and time lags of other AGNs \citep{Edelson2019}, microlensing observations \citep{Morgan2010}, 
and other more sparse AGN light curves \citep{Kelly2009,MacLeod2010}. It could also be interesting to 
apply our CHAR model to fit the extremely short-timescale ($\gtrsim 100$ Hz) variability observed in 
black hole X-ray binaries.

\acknowledgments 
We thank the anonymous referee for his/her helpful comments that improved the manuscript. 
We thank Krista Lynne Smith and Vishal P. Kasliwal for the useful discussions of the \textit{Kepler} data. 
We thank Bifang Liu for beneficial discussion. 
MYS acknowledges support from the National Natural Science Foundation of China (NSFC-11603022, 
11973002). 
MYS and YQX acknowledge support from the 973 Program (2015CB857004), the National 
Natural Science Foundation of China (NSFC-11890693, 11421303), the CAS Frontier Science Key 
Research Program (QYZDJ-SSW-SLH006), and the K.C. Wong Education Foundation. 
WNB acknowledges support from NSF grant AST-1516784 and NASA grant 80NSSC19K0961. 
WMG acknowledges support from the National Natural Science Foundation of China (NSFC-11925301). 
JRT acknowledges support from NASA STScI grants HST-GO-15260 and HST-GO-15650. 
ZYC acknowledges support from the National Natural Science Foundation of China (NSFC-11873045). 
TL acknowledges support from the National Natural Science Foundation of China (NSFC-11822304).

\software{AstroML \citep{astroml}, emcee \citep{emcee}, Matplotlib \citep{Hunter2007}, Numpy \& Scipy \citep{scipy}, 
PYCCF \citep{PYCCF}.}

\end{document}